\documentclass[12pt,a4paper]{article}
\usepackage[usenames,dvipsnames]{xcolor}
\usepackage{jheppub}  
\usepackage{amssymb} 
\usepackage{amsmath}
\usepackage{amsthm}
\usepackage{mathtools}
\usepackage{amsfonts}    
\usepackage{dsfont}
\usepackage{pdfpages}
\usepackage{verbatim}
\usepackage{tensor}
\usepackage{mathrsfs}
\usepackage{textgreek}
\usepackage{braket}
\usepackage{tikz}
\usetikzlibrary{decorations.pathreplacing, decorations.markings,calc,shapes.misc,decorations.pathmorphing,patterns.meta, math}
\usepackage{listings}
\usepackage[most]{tcolorbox}

\definecolor{codegreen}{rgb}{0,0.6,0}
\definecolor{codegray}{rgb}{0.5,0.5,0.5}
\definecolor{codepurple}{rgb}{0.58,0,0.82}
\definecolor{backcolour}{rgb}{0.95,0.95,0.92}

\lstdefinestyle{mystyle}{
    backgroundcolor=\color{backcolour},   
    commentstyle=\color{codegreen},
    keywordstyle=\color{magenta},
    numberstyle=\tiny\color{codegray},
    stringstyle=\color{codepurple},
    basicstyle=\ttfamily\footnotesize,
    breakatwhitespace=false,         
    breaklines=true,                 
    captionpos=b,                    
    keepspaces=true,                 
    numbers=left,                    
    numbersep=5pt,                  
    showspaces=false,                
    showstringspaces=false,
    showtabs=false,                  
    tabsize=2
}

\newcommand{\ba}{\begin{align}}

\newcommand{\be}{\begin{equation}}
\newcommand{\ee}{\end{equation}}
\def\bd{\begin{tikzpicture}}
\def\ed{\end{tikzpicture}}

\newcommand\Res{\mathop{\text{Res}}}
\allowdisplaybreaks[1]

\newcommand\SL{\text{SL}}

\newcommand\CC{\mathbb{C}}
\newcommand\ZZ{\mathbb{Z}}
\newcommand\RR{\mathbb{R}}

\newcommand\CP{\mathbb{CP}}

\hypersetup{
    pdfencoding=unicode,
	colorlinks=true,
	urlcolor=Maroon,
	linkcolor=RoyalBlue,
	citecolor=Maroon,
	pdftitle={A localising \texorpdfstring{$\mathrm{AdS}_3$}{AdS3} sigma model},
	pdfauthor={Lorenz Eberhardt and Matthias R.~Gaberdiel},
	pdfdisplaydoctitle=true,
	pdfstartview=FitH,
	linktocpage=true
}

\title{A localising AdS$_3$ sigma model}

\author[1]{Lorenz Eberhardt}\emailAdd{l.eberhardt@uva.nl}
\author[2,3]{\!\!, Matthias R.~Gaberdiel}\emailAdd{gaberdiel@itp.phys.ethz.ch}
\affiliation[1]{Institute for Theoretical Physics, University of Amsterdam, PO Box 94485, 1090 GL Amsterdam, The Netherlands}
\affiliation[2]{Institut f\"ur Theoretische Physik, ETH Zurich,
CH-8093 Z\"urich, Switzerland 
}
\affiliation[3]{Kavli Institute for Theoretical Sciences, University of Chinese Academy of Sciences,\\
Beijing 100190, China}

\abstract{We construct a CFT with $\mathfrak{sl}(2,\RR)_k$ symmetry at the `tensionless' point $k=3$, which is \emph{distinct} from the usual $\mathrm{SL}(2,\mathbb{R})_{k=3}$ WZW model. This new CFT is much simpler than the generic WZW model: in particular its three-point functions feature momentum-conserving delta functions, and its higher-point functions localise to covering map configurations in moduli space. We establish the consistency of the theory by explicitly deriving the four-point function from the three-point data via a sum over conformal blocks. The main motivation for our construction comes from holography, and we show that various simple supersymmetric holographic dualities for $k_{\rm s}=1$ ($k=3$) can be constructed by replacing the $\mathrm{AdS}_3$ factor on the worldsheet with this alternative theory. This includes in particular the prototypical case of $\mathrm{AdS}_3 \times \mathrm{S}^3 \times \mathbb{T}^4$, as well as the recently discussed example of $\mathrm{AdS}_3 \times \mathrm{S}^3 \times \mathrm{S}^3 \times \mathrm{S}^1$. However, our analysis does not require supersymmetry and also applies to bosonic ${\rm AdS}_3$ backgrounds (at $k=3$). 
}

\begin{document}

\maketitle

\makeatletter
\g@addto@macro\bfseries{\boldmath}
\makeatother

\section{Introduction}
String theory on $\text{AdS}_3$ with pure NS-NS flux has provided one of the most fruitful arenas to explore the AdS/CFT correspondence. The worldsheet theory is under unparalleled control and allows for exact (in $\alpha'$) comparisons between string theory and the dual CFT.
It was furthermore realised in \cite{Gaberdiel:2018rqv, Eberhardt:2018ouy, Eberhardt:2019ywk} that string theory on the background $\mathrm{AdS}_3 \times \mathrm{S}^3 \times \mathbb{T}^4$ drastically simplifies further in the tensionless limit, with one unit of NS-NS flux, where it provides the string theory dual of the symmetric orbifold of $\mathbb{T}^4$. The traditional worldsheet description in the RNS-formalism of $\mathrm{AdS}_3$ backgrounds involves the $\mathrm{SL}(2,\mathbb{R})$ WZW model, which is an exactly solvable, albeit relatively complicated 2d CFT \cite{Teschner:1997ft, Teschner:1999ug, Maldacena:2001km, Dei:2021xgh, Dei:2021yom}. This description partially breaks down in the tensionless limit and the equivalence of the spectrum with the symmetric orbifold of $\mathbb{T}^4$ was understood in \cite{Eberhardt:2018ouy} using the so-called hybrid formalism \cite{Berkovits:1999im}.

This duality was later extended to backgrounds with an arbitrary amount of NS-NS flux, for which the worldsheet theory is still completely solvable, at least for simple enough correlators. Through direct computation, a dual CFT was proposed in \cite{Eberhardt:2021vsx}, see also \cite{Eberhardt:2019qcl, Balthazar:2021xeh, Dei:2022pkr, Hikida:2023jyc}. For bosonic strings on $\mathrm{AdS}_3 \times {\rm X}$, where we take ${\rm X}$ to be described by a unitary CFT of central charge $c_{\rm X}=26-\frac{3k}{k-2}$, the dual CFT is given by a deformation of a symmetric orbifold of $\mathbb{R}_Q \times {\rm X}$. Here, $\mathbb{R}_Q$ is a linear dilaton factor with slope $Q=\frac{k-3}{\sqrt{k-2}}$, so that $\mathbb{R}_Q \times {\rm X}$ has central charge $6k$. The marginal operator that has to be turned on is a certain non-normalizable operator from the twist-2 sector of the symmetric orbifold. Thus, the CFT dual to a general $\mathrm{AdS}_3 \times {\rm X}$ background is somewhat analogous to Liouville theory as constructed by deforming a linear dilaton theory by the operator $\mathrm{e}^{2b \phi}$. This correspondence has also been generalised and tested for the superstring on various backgrounds \cite{Eberhardt:2021vsx, Yu:2024kxr, Yu:2025qnw}.
\medskip

For $k=3$ ($k_\text{s}=1$ in the superstring),\footnote{We shall use the convention that the levels of the ${\cal N}=1$ supersymmetric affine algebras are denoted by $k_{\rm s}$, while those of the (decoupled) bosonic algebras are denoted by $k$. For $\mathfrak{sl}(2,\RR)$ the two levels are related as $k=k_{\rm s}+2$, whereas for $\mathfrak{su}(2)$ we have instead $k=k_{\rm s}-2$.} the slope of the dilaton vanishes, meaning that the effective string coupling is constant throughout $\mathrm{AdS}_3$. This phenomenon has appeared before in the literature in various guises \cite{Seiberg:1999xz, Giveon:2005mi}. In particular, for $k=3$ the symmetric orbifold $\text{Sym}^N(\RR \times {\rm X})$ is a consistent CFT \emph{without} the need to turn on the marignal operator. In turn, this would imply that there are must be \emph{two} consistent CFTs with $\mathfrak{sl}(2,\mathbb{R})_k$ current symmetry at $k=3$: the $\mathrm{SL}(2,\mathbb{R})_{k=3}$ theory that is the restriction of the generic case discussed in detail in \cite{Dei:2021xgh,Dei:2021yom,Dei:2022pkr} to $k=3$; and a special second solution that only exists at $k=3$, and that we shall denote by $\mathrm{SL}(2,\mathbb{R})_{k=3}'$.\footnote{More precisely, we mean the WZW model based on the universal cover of $\SL(2,\RR)$, but we will suppress this in our notation.}  The latter model has correlators that are delta-function localised to the configurations that admit a covering map, whereas for the former theory this is not the case, see e.g.\  eq.~(\ref{eq:three point function integral expression}) below. Correspondingly, for $k=3$, there must also be two different string backgrounds that one may legitimately call $\mathrm{AdS}_3 \times {\rm X}$.

The main point of this paper is to demonstrate that such a consistent second theory indeed exists. Given that the dual CFT is much simpler, the worldsheet theory will also be much simpler. In particular, it will only feature principal series representations, and the structure constants vanish unless the $\mathfrak{sl}(2,\RR)$ spins satisfy a condition that translates to momentum conservation in the free boson factor of the dual CFT. Higher-point functions exhibit the localisation phenomenon that was first discussed in \cite{Eberhardt:2019ywk}: their correlation functions have delta-function support in the moduli space of surfaces, which reduces the integral over moduli space to a discrete sum, thereby making contact with the sum over covering surfaces that is encountered in the dual CFT.

The fact that only the principal series representations appear in the worldsheet spectrum suggests that 
the CFT that we are discussing here describes a \emph{Euclidean} $\mathrm{AdS}_3$ target space. Strictly speaking, its symmetry algebra is thus $\mathfrak{sl}(2,\CC)$. To properly describe it, one has to employ the so-called $y$-basis for its correlators \cite{Dei:2021xgh, Dei:2021yom}. However, for ease of presentation, we will initially ignore this subtlety in the reality conditions of the theory, and discuss it only at the end of Section~\ref{sec:crossing symmetry}.
\medskip

Our findings clear up a number of confusions in the literature. First of all, they suggest that it should after all be possible to formulate the duality between string theory on ${\rm AdS}_3 \times {\rm S}^3 \times \mathbb{T}^4$ with one unit $k_{\rm s}=1$ of NS-NS flux and the symmetric orbifold of $\mathbb{T}^4$ in the NS-R formalism. On the face of it this seems impossible since at $k_{\rm s}=1$ the $\mathfrak{su}(2)^{(1)}$ factor, describing the propagation on ${\rm S}^3$, then also has $k_{\rm s}=1$, and as a consequence the (decoupled) bosonic $\mathfrak{su}(2)$ factor has negative level  $k=-1$. Because of this issue, the analysis of \cite{Eberhardt:2018ouy} was done using the somewhat more involved hybrid formalism of \cite{Berkovits:1999im} where this problem does not arise (since the worldsheet fermions sit in spinor representations). However, the localisation analysis of \cite{Eberhardt:2019ywk} was originally done for the bosonic $\mathfrak{sl}(2,\RR)_k$ algebra at $k=3$ --- the hybrid version of this argument was only later found in \cite{Dei:2020zui} --- and one may suspect that the AdS/CFT duality should also work for a suitable NS-R background with supersymmetric level $k_s=1$ (or bosonic level $k=3$) for the $\mathfrak{sl}(2,\RR)^{(1)}$ factor. We shall sketch how this may work for the case of ${\rm AdS}_3 \times {\rm S}^3 \times \mathbb{T}^4$ in Section~\ref{sec:T4}.

Another motivation for this paper came from the recent concrete proposal for the 
special case of ${\rm AdS}_3 \times {\rm S}^3 \times {\rm S}^3 \times {\rm S}^1$ with $k_{\rm s}=1$  \cite{Gaberdiel:2024dva}, see also \cite{Gaberdiel:2018rqv,Giribet:2018ada} for earlier related work. For that case, the supersymmetric levels  of the two $\mathfrak{su}(2)^{(\pm)}$ algebras are $k_{\rm s}^{(\pm)}=2$, and thus the corresponding bosonic levels are $k^{(\pm)}=0$, which does not cause any inconsistency in the NS-R description. The analysis of \cite{Gaberdiel:2024dva} suggested that there should be an exact duality to an undeformed symmetric orbifold theory. However, as was already mentioned there, this requires the existence of the localising second solution,\footnote{Because of this issue, some questions were raised about the proposal of  \cite{Gaberdiel:2024dva} in \cite{Chakraborty:2025nlb}.} and if we take the tensionless AdS$_3$ string to be described by this theory, the analysis of this paper effectively prove the duality of \cite{Gaberdiel:2024dva} (at least in the planar limit). 

While it is satisfying to clear up these special cases, our arguments actually suggest much more: the same arguments apply to any NS-R background that leads to a bosonic $\mathfrak{sl}(2,\RR)_k$ theory at level $k=3$, and even supersymmetry does not seem to be required, i.e.\ they predict that  bosonic string theory on ${\rm AdS}_3 \times {\rm X}$, where the AdS factor is described by the second $\mathfrak{sl}(2,\RR)_k$ theory at level $k=3$, is dual to the symmetric orbifold of $(\RR \times {\rm X})$.\footnote{Note that the central charge works out as expected: for a critical bosonic string the ${\rm X}$ factor must have $c=17$ so that the total worldsheet theory has $c=26$. The seed theory of the symmetric orbifold should then be $c=17+1=18$ since the non-compact boson of the $\mathfrak{sl}(2,\RR)_3$ factor survives (in addition to the excitations from ${\rm X}$). This then fits together with the form of the symmetric orbifold correlators, see eq.~(\ref{eq:symorb}).} Given that these symmetric orbifold theories are consistent and stable, the corresponding bosonic string is actually tachyon-free -- the technical reason for this is that our worldsheet theory only involves states with spectral flow $w\geq 1$, and thus the would-be tachyon is not part of the spectrum. However, unlike other tachyon-free bosonic string theories, our model has a highly non-trivial spectrum. 
\medskip

The paper is organised as follows. In Section~\ref{sec:CFT} we motivate our ansatz for the correlation functions of the localising $\mathfrak{sl}(2,\RR)$ theory at $k=3$. The core part of the paper is Section~\ref{sec:crossing symmetry} where we show that these three-functions lead to crossing symmetric  four-point functions. We first do this in the usual basis of vertex operators that partially ignores issues stemming from the Euclidean nature of the model, but we also demonstrate in Section~\ref{subsec:Euclidean theory} how the arguments work (more properly) in the Euclidean setting. We also give a general argument based on the explicit form of the correlators in the $y$-basis of \cite{Dei:2021yom,Dei:2022pkr} in Section~\ref{subsec:general argument}. We discuss various implications of the existence of this localising solution for AdS holography in Section~\ref{sec:holography}. Section~\ref{sec:concl} contains our conclusions and suggests various open problems. We spell out our conventions in detail in Appendix~\ref{app:Conventions}.

\section{A crossing symmetric CFT} \label{sec:CFT}

The main aim of our paper is to define a new crossing symmetric CFT based on the WZW model of $\mathfrak{sl}(2,\RR)_k$  for the special case $k=3$.\footnote{Since the theory will only consist of the continuous representations and possesses the reality conditions of a Euclidean signature target space, it is more correct to think of it as corresponding to ${\rm SL}(2,\CC)/{\rm SU}(2)$, and we shall comment on this below, see Section~\ref{subsec:Euclidean theory}. While our theory is only interesting for $k=3$, many aspects of it can be understood for generic $k$, and we shall therefore phrase many statements for the general case.} This theory differs from the generic $k$ solution of crossing symmetry found in \cite{Dei:2021xgh} in that the worldsheet correlators are delta-function localised on the configurations that have an interpretation in terms of covering maps. In order to motivate our proposal, let us first review briefly the structure of the solution for general $k$.

\subsection{The spectrum and structure constants}

The theory we are interested in consists of the states in the principal series representaitons of $\mathfrak{sl}(2,\mathbb{R})$ with $\mathfrak{sl}(2,\mathbb{R})$ spin $j=\frac{1}{2}+i p$, but does not involve any states from the  discrete representations. As is familiar from the usual discussions of string theory on ${\rm AdS}_3$ \cite{Maldacena:2000hw}, in addition to these highest weight representations, we also consider the images under spectral flow by $w$ units with $w \in \ZZ_{\ge 1}$. The vertex operators of our theory can therefore be labelled by $V_{j,h,\bar{h}}^w(x;z)$ where the spin $j$ is determined by $j=\frac{1}{2}+i p$ through the momentum $p$, $w$ is the amount of spectral flow, and $z$ denotes the insertion point on the `worldsheet' (where our $2d$ CFT lives), while $x\in\mathbb{CP}^1$ is an internal coordinate that can be identified with the position on the boundary of ${\rm AdS}_3$. Furthermore, $h$ and $\bar{h}$ are the eigenvalues with respect to the left- and right-moving Cartan generator $J^3_0$ and $\bar{J}^3_0$ of $\mathfrak{sl}(2,\RR)$, i.e.\ the `magnetic quantum numbers', of the state at $x=0$, see e.g.\  \cite{Eberhardt:2019ywk,Dei:2021xgh} for a more detailed explanation of our conventions. The vertex operators are characterised by the OPEs with the symmetry currents, and these are spelled out explicitly in Appendix~\ref{app:Conventions}. For most of the paper we shall concentrate on the case where $z\in\mathbb{CP}^1$, i.e.\ we shall only consider the CFT on the sphere. 

For generic $k$, the three-point function in the $\mathfrak{sl}(2,\mathbb{R})$ WZW model is only non-zero provided that the triangle inequalities $w_i \leq w_j+w_\ell +1$ with $\{i,j,\ell\}$ all distinct, are satisfied. It depends significantly on the parity of $\sum_i w_i$, and was explicitly determined (by systematically imposing Ward identities and crossing symmetry) in \cite[eq.~(5.23)]{Dei:2021xgh}, see also \cite{Bufalini:2022toj}. For $\sum_i w_i \in 2 \ZZ+1$ and $w_i \le w_j+w_\ell-1$ with $\{i,j,\ell\}$ distinct,\footnote{There are similar expressions for $\sum_i w_i \in 2\ZZ$ or for the extremal case $w_i=w_j+w_\ell+1$ for some $\{i,j,\ell\}$ with $\sum_i w_i \in 2\ZZ+1$, but it does not have a pole for $j_1+j_2+j_3=\frac{k}{2}$ that will be important below.} the three-point function takes the form\footnote{The notation $|\bullet|^2$ means that we take the product with the right-moving analogue. This only complex conjugates the $t_i$'s, but replaces $h_i \to \bar{h}_i$ and does not complex conjugate the exponents. We also use the opposite conventions to \cite{Dei:2021xgh} and replaced $j_i \to 1-j_i$. This does not matter for the principal series representations that we discuss in this paper and leads to more convenient expressions.}
\begin{align}
&\left\langle V_{j_1,h_1,\bar{h}_1}(0;0)V_{j_2,h_2,\bar{h}_2}(1;1)V_{j_3,h_3,\bar{h}_3}(\infty;\infty) \right\rangle\nonumber\\
&\qquad=\tilde{D}(j_1,j_2,j_3) N^{k -6+2 (j_1+j_2+j_3)} \Big|\prod_{i=1}^3 a_i^{\frac{k}{4}(w_i-1)-h_i} w_i^{-\frac{k}{4}(w_i+1)+1-j_i} \Pi^{-\frac{k}{2}}\Big|^2\nonumber\\
&\qquad\qquad\times \int \prod_{i=1}^3 \mathrm{d}^2 t_i \ \Bigg| \prod_{i=1}^3 t_i^{-\frac{k w_i}{2}+h_i-j_i} (1-t_i)^{\frac{k w_i}{2}-h_i-j_i}\nonumber\\
&\hspace{5cm}\times\left(1-\frac{w_1 t_1}{N}-\frac{w_2 t_2}{N}-\frac{w_3 t_3}{N}\right)^{\frac{k}{2} -3+ j_1+j_2+j_3}\Bigg|^2\ . \label{eq:three point function integral expression}
\end{align}
For this choice of spectral flows $w_i$, there is always a corresponding covering map
\be 
\gamma: \CP^1 \to \CP^1\ ,
\ee
ramified over the three points $z_i$ with ramification indices $w_i$, such that $\gamma(z_i)=x_i$.
Except for the prefactor $\mathcal{N}(j_1)D(\frac{k}{2}-j_1,j_2,j_3)$ that we are about to discuss, all other parameters are defined in terms of the corresponding covering map: in particular $N=\frac{1}{2}(w_1+w_2+w_3-1)$ is the degree of the covering map, while $a_i$ are the coefficients appearing in (\ref{eq:covering map coefficients}), and $\Pi$ is the product over the $N$ residues, see eq.~(\ref{eq:Pi definition}). 

The coefficient 
\be 
\tilde{D}(j_1,j_2,j_3)=\mathcal{N}(1-j_1)D(\tfrac{k}{2}-1+j_1,1-j_2,1-j_3)
\ee
is related to the structure constant $D(j_1,j_2,j_3)$ of the unflowed correlators, see  \cite[eq.~(5.36) and below]{Maldacena:2001km}, and takes a similar form as the DOZZ structure constants of Liouville theory \cite{Teschner:1997ft}, and $\mathcal{N}(1-j_1)$ is a coefficient needed to make the expression symmetric in the three operators; the detailed expressions are for example given in \cite[Appendix~B]{Dei:2021xgh}. Generically, the structure constants \eqref{eq:three point function integral expression} are non-zero, but have a series of poles in the spin; these are studied in detail in \cite{Dei:2022pkr}, see also \cite{Maldacena:2001km,Aharony:2004xn} for earlier work. The first and simplest of these poles occurs for 
\be\label{eq:pole}
j_1+j_2+j_3=\frac{k}{2} \quad \Longleftrightarrow \quad i \bigl(p_1+p_2+p_3 \bigr)= \frac{k-3}{2}  \ . 
\ee
Thus, for $k=3$, and only for that value, the simple pole can arise for the physical values $p_i\in\RR$. Similar to the Coulomb gas construction of Liouville theory, the three-point functions simplify dramatically when we take the residues at this value. Indeed, we have
\begin{align}
    \!\Res_{j_1+j_2+j_3=\frac{k}{2}}\!\left\langle V_{j_1,h_1,\bar{h}_1}(0;0)V_{j_2,h_2,\bar{h}_2}(1;1)V_{j_3,h_3,\bar{h}_3}(\infty;\infty) \right\rangle=\Big|\hat{\Pi}^{-\frac{k}{2}} \prod_{i=1}^3  a_i^{\frac{k}{4}(w_i-1)-h_i} \Big|^2 , 
\end{align}
where, in order to streamline notation, we have defined the convenient combination
\be\label{eq:hatPi}
\hat{\Pi} = \prod_i w_i^{\frac{1}{2}(w_i+1)} \, \Pi \ ,
\ee
see also Appendix~\ref{app:Conventions}.
The idea of our proposal is to take the structure constants of our new theory to simply be the residues of these poles and impose in addition momentum conservation. For $k=3$, we can then parametrise our vertex operators more naturally through their momentum $p_i$ and write $V_{p,h,\bar{h}}^{w}(x;z)$. We thus set
\begin{multline} 
 \left\langle V_{p_1,h_1,\bar{h}_1}^{w_1}(0;0)V_{p_2,h_2,\bar{h}_2}^{w_2}(1;1)V_{p_3,h_3,\bar{h}_3}^{w_3}(\infty;\infty) \right\rangle_{\rm new} \\
 =\Bigl|\hat{\Pi}^{-\frac{3}{2}}\, \prod_{i=1}^3   a_i^{\frac{3}{4}(w_i-1)-h_i} \Bigr|^2\   \delta(p_1+p_2+p_3)\ .\label{eq:proposal 3pt function}
\end{multline}
One can nicely generalise this to arbitrary $x_i$ and $z_i$ by taking the $\hat{\Pi}$ and $a_i$ to be those of the appropriate covering map, see Appendix~\ref{subapp:Conventionscoveringmaps} for the precise definitions, leading to 
\begin{align}
    \left\langle \prod_{i=1}^3 V_{p_i,h_i,\bar{h}_i}(x_i;z_i)\right\rangle_{\rm new}=\prod_{i<j} |z_i-z_j|^{4p_i p_j} \Bigl|\hat\Pi^{-\frac{3}{2}}\prod_{i=1}^3 a_i^{\frac{3(w_i-1)}{4}-h_i}\Bigr|^2 \delta(p_1+p_2+p_3) \ . \label{eq:three-point function}
\end{align}
Since we assume here the existence of a covering map, it is understood that these three-point functions are only non-zero provided that the triangle inequalities on the $w_i$ and the parity constraint $\sum_i w_i \in 2\ZZ+1$ is satisfied.

\subsection{Higher point functions and localisation}

For the four-point functions a similar result was worked out in \cite[eq. (3.29)]{Dei:2022pkr}. It was shown there that the four-point function also simplifies dramatically for $k=3$ when $\sum_i p_i=0$. More specifically, it has a singularity of the form $|z-z_\gamma|^{-2}$, where $z_\gamma$ is the value of the cross-ratio $z$ for which a covering map with the appropriate branchings exists. As a consequence the string correlator (where we integrate over the cross ratio $z$) diverges, and the string four-point function contains a residue at $\sum_i p_i=0$. The idea is to replace this divergence with a delta function (for $\sum_i p_i=0$). We thus propose the modified four-point function
\begin{multline}
\left\langle V_{p_1, h_1,\bar{h}_1}^{w_1}(0;0) V_{p_2, h_2,\bar{h}_2}^{w_2}(1;1) V_{p_3, h_3,\bar{h}_3}^{w_3}(\infty;\infty) V_{p_4, h_4,\bar{h}_4}^{w_4}(x;z) \right\rangle_{\rm new} \\
=  \sum_{\gamma} \delta^{(2)}(z-z_\gamma)\,   |z|^{4p_1p_4} \, |1-z|^{4p_2p_4}\Big|\hat\Pi^{-\frac{3}{2}} \, 
 \prod_{i=1}^4  a_i^{\frac{3(w_i-1)}{4}-h_i}\Big|^2 \delta\Big(\sum\nolimits_i p_i \Big) \ . \label{eq:four-point function0} 
\end{multline}
This also generalises nicely to arbitrary $x_i$ and $z_i$, for which we get
\begin{multline}
    \left\langle \prod_{i=1}^4 V_{p_i, h_i,\bar{h}_i}^{w_i}(x_i;z_i) \right\rangle_{\rm new}=\sum_{\gamma} \delta^{(2)}(z-z_\gamma)\,   \prod_{i<j}|z_i-z_j|^{4p_ip_j}\delta\Big(\sum\nolimits_i p_i \Big)\\ \times \Big|\hat\Pi^{-\frac{3}{2}} 
 \prod_{i=1}^4  a_i^{\frac{3(w_i-1)}{4}-h_i}\Big|^2  \ ,  \label{eq:four-point function}
\end{multline}
where, as before, the $a_i$ and $\hat\Pi$, see eq.~(\ref{eq:hatPi}), are the parameters of the relevant covering map. We note that the answer is the product of a free boson correlator 
\be
\prod_{i<j}|z_i-z_j|^{4p_ip_j}\delta\Big(\sum\nolimits_i p_i \Big) \ , 
\ee
together with the coefficients one would expect for the symmetric orbifold correlator of twist fields for which the central charge of the seed theory is $c=18$ \cite{Dei:2019iym},
\be\label{eq:symorb}
\sum_{\gamma} \delta^{(2)}(z-z_\gamma)\,\Big|\hat\Pi^{-\frac{3}{2}} \,  \prod_{i=1}^4  a_i^{\frac{3(w_i-1)}{4}-h_i}\Big|^2\ .
\ee
In fact, this then also immediately suggests a generalisation to arbitrary $n$-point functions, 
\begin{multline}
    \left\langle \prod_{i=1}^n V_{p_i, h_i,\bar{h}_i}^{w_i}(x_i;z_i) \right\rangle_{\rm new}=\sum_{\gamma} \prod_{i=4}^{n}\delta^{(2)}(z_i-z_i^\gamma)\,   \prod_{i<j}|z_i-z_j|^{4p_ip_j}\delta\Big(\sum\nolimits_i p_i \Big)\\ \times \Big|\hat\Pi^{-\frac{3}{2}} 
 \prod_{i=1}^n  a_i^{\frac{3(w_i-1)}{4}-h_i}\Big|^2  \ ,  \label{eq:proposal npt function}
\end{multline}
where the $z_i$ need to satisfy the $n-3$ constraints $z_i=z_i^\gamma$ in order for a covering map (with these branching indices) to exist. 

\subsection{The identity and the two-point function}
As for the ordinary $\SL(2,\RR)$ WZW model, the theory that we consider here does not have an identity field.\footnote{From a holographic point of view, it however has a special field that corresponds to the identity field in spacetime. It is given by the operator $V^{w=1}_{p=0,h=\bar{h}=0}$, which has worldsheet conformal weight $\Delta=1$ and hence satisfies the mass-shell condition. It is the dilaton zero mode and behaves anlogously to the marginal operator $\mathrm{e}^{2b \phi}$ in Liouville theory \cite{Giveon:2001up,Kim:2015gak}.} This means that we cannot obtain the two-point function normalisation from the three-point function normalisation in a simple way. This is a situation similar to what one encounters e.g.\ in timelike Liouville theory \cite{Harlow:2011ny, Ribault:2015sxa}. Instead, we will simply specify a two-point function normalisation that leads to a crossing symmetric four-point function,
\be 
\langle V_{j,h,\bar{h}}^{w}(0;0) V_{j',h',\bar{h}'}^{w'}(\infty;\infty) \rangle=\frac{1}{w} \, \delta^{(2)}(h-h') \, \delta(j+j'-1) \, \delta_{w,w'}\ , \label{eq:two-point normalization}
\ee
where $\delta^{(2)}(h)=\frac{1}{2}\delta_{h-\bar{h}}\, \delta(h+\bar{h})$. More precisely, this two-point normalisation should be interpreted in the Euclidean theory, see Section~\ref{subsec:Euclidean theory}.

\section{Conformal block expansion of the four-point function} \label{sec:crossing symmetry}
The goal of this section is to verify that the four-point function \eqref{eq:four-point function} is reproduced by the conformal block expansion from the three-point functions \eqref{eq:three-point function}. 
This will then in particular ensure crossing symmetry of the model. 
We will actually be somewhat more general and keep $k$ arbitrary since our argument works for any $k$, as long as the momentum conservation for an $n$-point function takes the form
\be 
i\sum_j p_j=\frac{(n-2)(k-3)}{2}\ .
\ee
As we mentioned above, momentum conservation within the physical spectrum can only be satisfied for $k=3$ and thus we only have a fully consistent CFT for that value. Special cases of the conformal block expansion and crossing symmetry have been checked in the generic $\mathrm{SL}(2,\mathbb{R})_k$ WZW model in \cite{Iguri:2024yhb}.

We will first perform the calculation by brute force, i.e.\  by expanding the conformal blocks to the first subleading order in a small cross ratio expansion, see Sections~\ref{sec:confblockexp} and \ref{sec:confblockexpsub}. The calculation is quite technical, and it requires various ingredients that we shall first derive, see Sections~\ref{sec:recursion}-\ref{subsec:Kac matrix}, and that are hopefully of independent value regardless of the specific application we have in mind. (In particular, we will explain a shorter route to derive the form of the three-point functions from the Ward identities than was originally discussed in \cite{Eberhardt:2019ywk}.)
The appearance of the delta function in \eqref{eq:four-point function} is at first somewhat formal, and we also explain how to do this more properly by going to the Euclidean theory, see Section~\ref{subsec:Euclidean theory}. Finally, we give a general argument for the consistency of the four-point function in Section~\ref{subsec:general argument}. 

\subsection{Rederiving the recursion relations}\label{sec:recursion}
To warm up, we start by rederiving the $h_i$-dependence of the three-point functions \eqref{eq:three-point function}. This was first done in \cite{Eberhardt:2019ywk} through somewhat involved computations that yielded a recursion relation on the three-point functions in $h_1$, $h_2$ and $h_3$. We will explain an arguably simpler route to derive these recursion relations that generalise nicely to the computation of three-point functions with a descendant.

\paragraph{Ward identities.} Usually, we obtain Ward identities in a correlation function by inserting the symmetry current $J^a(z)$ into a correlator, evaluating all the residues as $z \to z_i$ and requiring that the sum of the residues vanishes. One can in principle modify this procedure by considering instead of the individual $J^a(z)$ suitable combinations $\sum_a f_a(z)J^a(z)$ for some meromorphic functions $f_a(z)$ whose only poles (on the Riemann sphere) are at $z=z_i$; then one can proceed as before, i.e.\ insert this combination into the correlator and require that the residues at $z=z_i$ sum to zero. This gives us relations between correlators with descendants, and hence allows us to reduce correlators with descendant fields to those of the corresponding primaries.

For a spectrally flowed correlator, we want to choose the functions $f_a(z)$ such that the residues only pick out the leading singularities of $J^a(z)$ with the three primary fields $V_{j_i,h_i,\bar{h}_i}^{w_i}(x_i;z_i)$, which we keep in general location $(x_i,z_i)$. The OPEs of the currents with the primary fields are given in \eqref{eq:x-OPEs} --- importantly they feature higher order poles, while the modes we want to pick out by taking the residues are those in \eqref{eq:leading J action}. This tells us that the three functions $f_+(z)$, $f_3(z)$ and $f_-(z)$ should behave as follows as $z \to z_i$,
\begin{subequations}
\begin{align}
    f_-(z) &\sim \mathcal{O}\bigl((z-z_i)^{-w_i}\bigr)\ , \\
    f_3(z)+2x_i f_-(z) &\sim \mathcal{O}(1)\ , \\
    f_+(z)+x_i f_3(z)+x_i^2 f_-(z) &\sim \mathcal{O}\bigl((z-z_i)^{w_i}\bigr)\ ,
\end{align} \label{eq:f recursion relation behaviour xi}%
\end{subequations}
since then
\begin{align} 
&\sum_a f_a(z) J^a(z) V_{j_i,h_i,\bar{h}_i}^{w_i}(x_i;z_i) \nonumber\\[-8pt]
&\qquad\sim (f_+(z)+x_i f_3(z)+x_i^2 f_-(z)) (z-z_i)^{-w_i-1}\big(J^+_{w_i}V_{j_i,h_i,\bar{h}_i}^{w_i}\big)(x_i;z_i) \nonumber\\
&\qquad\qquad+(f_3(z)+2x_i f_-(z))(z-z_i)^{-1}\big(J^3_0V_{j_i,h_i,\bar{h}_i}^{w_i}\big)(x_i;z_i) \nonumber\\
&\qquad\qquad+ f_-(z) (z-z_i)^{w_i-1} \big(J^-_{-w_i}V_{j_i,h_i,\bar{h}_i}^{w_i}\big)(x_i;z_i) \label{eq:OPE spectrally flowed current combination}
\end{align}
only has first order poles, and the residue only involves the action of the spectrally flowed zero modes on the primary fields.

\paragraph{Constraining the functions $f_a$.} We can replace $x_i$ in \eqref{eq:f recursion relation behaviour xi} by the covering map $\gamma(z)$, since $\gamma(z)-x_i= \mathcal{O}((z-z_i)^{w_i})$ and the correction term does not modify the above behaviour. Let us hence denote
\begin{subequations}
\begin{align}
    F_-(z)&=f_-(z)\sim  \mathcal{O}((z-z_i)^{-w_i})\ , \\
    F_3(z)&=f_3(z)+2 \gamma(z) f_-(z)\sim \mathcal{O}(1)\ , \\
    F_+(z)&=f_+(z)+\gamma(z) f_3(z)+\gamma(z)^2 f_-(z) \sim \mathcal{O}((z-z_i)^{w_i})\ .
\end{align} \label{eq:F through f}%
\end{subequations}
$F_+(z)$ thus has a $w_i$-fold zero at all the insertion points of the vertex operators, while it potentially has double poles at the poles $\ell_a$ of the covering map. However, notice that the number of zeros $Z$ of $F_+$ counted with their multiplicity is greater than the number of poles $P$, 
\be 
Z \ge w_1+w_2+w_3 >w_1+w_2+w_3-1=2N\ge P\ ,
\ee
where we have used that the degree $N$ of the covering map is related to the ramification index by the Riemann Hurwitz formula, see eq.~\eqref{eq:Riemann Hurwitz}. $Z>P$ is impossible for a non-zero analytic function and thus $F_+(z) \equiv 0$.

We can similarly constrain $F_3(z)$: writing 
\be
0 \equiv F_+(z) = f_+(z) + \gamma(z) \bigl( f_3(z) + \gamma(z) f_-(z) \bigr)  \ , 
\ee
the holomorphicity of $f_+(z)$ and $f_3(z)$ near $z=\ell_a$ implies that $\gamma(z) f_-(z) $ remains regular as $z \to \ell_a$, and thus $F_3(z)$ does not have any poles at $z=\ell_a$, or indeed anywhere else (including at $z=\infty$). By Liouville's theorem $F_3$ is constant, $F_3(z) \equiv c_1$ for some constant $c_1$.

Finally, we have to determine $F_-(z)$. Besides having at most $w_i$-fold poles at $z \to z_i$, it is also constrained by demanding that $f_3(z)$ and $f_+(z)$ remain regular as $z \to \ell_a$ (which are expressed through $F_-(z)$ by the conditions $F_3(z)= c_1$ and $F_+(z)=0$). This tells us that
\be 
F_-(z) \sim \frac{c_1}{\Pi_a} (z-\ell_a)+\mathcal{O}\bigl((z-\ell_a)^2\bigr) \qquad 
\hbox{as $z \to \ell_a\ $,}\label{eq:Fm first derivative constraint}
\ee
where $\Pi_a$ is the residue of the covering map at $z=\ell_a$, see eq.~\eqref{eq:Pi definition}. The most general solution to these constraints is given by
\be 
F_-(z)=\frac{c_1\partial^2 \gamma(z)}{ 2(\partial \gamma(z))^2}+\frac{c_1 z^2+c_2 z+c_3}{(z-z_1)(z-z_2)(z-z_3)\partial \gamma(z)} \ ,
\ee
where $c_2$ and $c_3$ are two new constants. Indeed, the first term precisely ensures that the constraint \eqref{eq:Fm first derivative constraint} is satisfied, while the second term vanishes to second order as $z \to \ell_a$. Its form is constrained by the fact that since $\partial \gamma(z) \sim \mathcal{O}((z-z_i)^{w_i-1})$, one can increase the pole order precisely by one with the explicit factors of $(z-z_i)^{-1}$. The numerator of the second term then has to be a polynomial with the coefficient of the $z^2$ term being fixed by requiring that $F_-(z)$ has no pole at $z=\infty$.

\paragraph{The recursion relations.} It is now simple to derive the Ward identities following from these choices by computing the residues of $\sum_a f_a(z) J^a(z)$. This gives us three equations. By choosing $c_1$, $c_2$ and $c_3$ appropriately, this leads to the three equations 
\begin{multline}\label{eq:recur}
a_\ell^{-1} d_{j_\ell}^-(h_\ell-\tfrac{k w_\ell}{2}) C(h_\ell-1)=\sum_{i=1}^3\Big(\frac{w_\ell}{N}-\delta_{i,\ell}\Big)a_i d_{j_i}^+(h_i-\tfrac{k w_i}{2}) C(h_i+1)\\ +\Big(2h_\ell-\frac{w_\ell}{N} \sum_{i=1}^3 h_i\Big) C(h_i)=0\ ,
\end{multline}
where $\ell=1,2,3$. Here $a_\ell$ are the coefficients appearing in the covering map, see eq.~(\ref{eq:covering map coefficients}), and $2N=w_1+w_2+w_3-1$ is the degree of the covering map. Furthermore, the parameters $d_j^\pm(m)$ describe our conventions for the action of the $\mathfrak{sl}(2,\RR)$ generators on the highest weight states, 
\be \label{eq:sl2action}
J_0^3 \ket{j,m}=m \ket{j,m}\ , \quad J_{0}^\pm \ket{j,m}= d_j^\pm(m) \ket{j,m\pm 1} \ , \quad 
d_j^\pm(m) = \big(m\pm j\big)  \ .
\ee
Furthermore, we use the shorthand notation $C(h_i+1)$ to denote the three-point function where the $i^{\rm th}$ conformal weight is shifted up by one unit, while the other weights are unmodified. It is not difficult to see that (\ref{eq:recur}) are, in fact, precisely the recursion relations derived in \cite{Eberhardt:2019ywk}.

\paragraph{The $j$-constraint.} A crucial observation in \cite{Eberhardt:2019ywk} was that these recursion relations are simple to solve when $\sum_i j_i=\frac{k}{2}$  --- in this case, we simply have
\be 
C(h_1,h_2,h_3)=C_0 \prod_{i=1}^3 a_i^{-h_i} \ , \label{eq:recursion solution}
\ee
where $C_0$ is independent of $h_i$, but can still depend on the other quantum numbers $j_i$ and $w_i$. The delta-function in \eqref{eq:three-point function} precisely implements the $j$-constraint $\sum_i j_i=\frac{k}{2}$ and thus this three-point function is compatible with the basic Ward identities.

\subsection{Three-point functions with a descendant}\label{sec:3ptdescendant}
We now want to compute the three-point functions with a level-1 descendant, i.e.\ the three-point functions
\be 
\big\langle (J_{\pm w_1-1}^\pm V_{j_1,h_1}^{w_1})(x_1;z_1) \prod_{i=2}^3 V_{j_i,h_i}^{w_i}(x_i;z_i) \big \rangle\ , \quad \big\langle (J_{-1}^3 V_{j_1,h_1}^{w_1})(x_1;z_1) \prod_{i=2}^3 V_{j_i,h_i}^{w_i}(x_i;z_i) \big \rangle\ . \label{eq:level-1 descendants}
\ee
We will need this when we compute the conformal blocks of the four-punctured sphere below.
To obtain them via Ward identities, we follow the same logic as above and insert the combination $\sum_a f_a(z) J^a(z)$ into a correlator of three primaries. To pick up the level-1 descendants \eqref{eq:level-1 descendants}, we relax our conditions on $f_a(z)$ at $z \to z_1$ and allow the pole order in \eqref{eq:f recursion relation behaviour xi} to be one higher. 

\paragraph{Constraining the functions $f_a$.} As before, it is convenient to work with the functions 
$F_a(z)$ that are obtained by replacing $x_i \to \gamma(z)$ as in \eqref{eq:F through f}, for which we now obtain the conditions
\begin{align}
    F_\pm(z) \sim \mathcal{O}((z-z_1)^{\pm w_1-1})\ , \quad F_3(z) \sim \mathcal{O}((z-z_1)^{-1})\ ,
\end{align}
while the conditions at $z_2$ and $z_3$ are the same as in \eqref{eq:F through f}. We can follow similar arguments as above to determine the most general solution to these constraints, and we find that it takes the form
\begin{subequations}
\begin{align}
    F_+(z)&=c_1 (z-z_2)(z-z_3) \partial \gamma(z)\ , \\
    F_3(z)&=c_1\, \frac{(z-z_2)(z-z_3)\partial^2 \gamma(z)}{\partial \gamma(z)}+2c_1 z+c_2+\frac{c_3}{z-z_1}\ , \\
    F_-(z)&=\frac{c_1(z-z_2)(z-z_3) \partial^3 \gamma(z)}{6 (\partial \gamma(z))^2}+\Big(2c_1 z+c_2+\frac{c_3}{z-z_1}\Big) \frac{\partial^2 \gamma(z)}{2(\partial \gamma(z))^2} \nonumber\\
    &\qquad+\frac{c_1 z^4+(c_2-2c_1 z_1) z^3+c_6 z^2+c_5 z+c_4}{(z-z_1)^2(z-z_2)(z-z_3)\partial \gamma(z)}\ , 
\end{align}
\end{subequations}
where we now have six free constants, $c_1$, \dots, $c_6$. Three of these solutions are those that lead to the recursion relations that we discussed above. 

\paragraph{Correlator with a descendant.} The remaining three recursion relations give a linear system of equations for the three descendant correlators \eqref{eq:level-1 descendants}. It is straightforward to solve this system of equations for the three unknowns, and assuming the $j$-constraint $\sum_i j_i=\frac{k}{2}$  is satisfied, we obtain 
\begin{multline}
  \frac{\big\langle J_{-1}^3 V_{j_1,h_1}^{w_1}(x_1;z_1) V_{j_2,h_2}^{w_2}(x_2;z_2) V_{j_3,h_3}^{w_3}(x_3;z_3) \big \rangle}{\big\langle V_{j_1,h_1}^{w_1}(x_1;z_1) V_{j_2,h_2}^{w_2}(x_2;z_2) V_{j_3,h_3}^{w_3}(x_3;z_3) \big \rangle} =\frac{b_1 (k ( w_1^2+2)-6 h_1 w_1+6 j_1)}{6 a_1 w_1^2}\\
    +\sum_{i=2}^3\frac{k (w_1^2+w_i^2-w_\ell^2+2)-6 j_i-6 (h_1 w_1+h_i w_i-h_\ell w_\ell)}{6
   w_1 (z_1-z_i)}\ , \label{eq:level-1 J3 correlator}
\end{multline}
where $\ell$ (as well as $i$) only run over the values $\{i,\ell\}=\{2,3\}$. Furthermore, we recall from \eqref{eq:covering map coefficients}, that $b_1$ is the sub-leading coefficient (relative to $a_1$) in the expansion of the covering map near $z_1$. The expressions for the other two descendants are similar. 

\subsection{Kac Matrix} \label{subsec:Kac matrix}
The next ingredient that is needed for the computation of the conformal block is the Kac matrix, that is the determinant of the matrix of inner products at some given worldsheet and target space conformal weight. It follows from (\ref{eq:sl2action}) that, in the state picture, the spectrally flowed highest weight state $\ket{j,h}^{(w)}$ satisfies (recall that $h = m + \frac{kw}{2}$, where $m$ is the eigenvalue of $J^3_0$ before spectral flow) 
\be \label{eq:specflowaction}
J_0^3 \ket{j,h}^{(w)}=h \ket{j,h}^{(w)}\ , \qquad J_{\pm w}^\pm\ket{j,h}^{(w)}=\big(h-\tfrac{kw}{2}\pm j\big)  \ket{j,h\pm 1}^{(w)}\ .
\ee
More positive modes annihilate the state. Similarly, more negative modes annihilate the corresponding bra state, which we take to have spin $1-j$ in order to have canonically normalised two-point functions,
\begin{align} 
\big(h-\tfrac{kw}{2}+j\big) {}^{(w)}\!\langle 1-j,h+1\, | \, j,h+1 \rangle^{(w)} &={}^{(w)}\!\bra{1-j,h+1}J_w^+ \ket{j,h}\\
&={}^{(w)}\!\bra{j,h}J_{-w}^- \ket{1-j,h+1}^{(w)} \\
&= \big(h-\tfrac{kw}{2}+j\big) {}^{(w)}\!\langle 1-j,h\, | \, j,h \rangle^{(w)}\ ,
\end{align}
i.e.\ the norm is independent of $h$. 

Similar to the familiar Virasoro case, we now want to compute the matrix of inner products at a given excitation level $N$ and spacetime conformal weight $h$. Since two-point functions vanish for different conformal weights, the Kac matrix will take a block diagonal form. A given block depends on the weight $h$ and the spectral flow $w$. However, since spectral flow is an automorphism, the Kac matrix actually only depends on $w$ via the combination $m=h-\frac{kw}{2}$. We can thus simply compute the Kac matrix in the unflowed sector. We will denote a given block at level $N$ by $M_{m}^{(N)}$ (where the other parameters $k$ and $j$ are left implicit). Thus for a choice of basis states $\ket{\psi_1},\dots,\ket{\psi_{P_3(N)}}$ in every block, we set\footnote{The size of the blocks is equal to the number of partitions with three different colors given by the generating function $\sum_{N \ge 0} P_3(N) q^N=\prod_{n=1}^\infty (1-q^n)^{-3}$.}
\be 
\big(M_m^{(N)}\big)_{i,l}= \langle \psi_i \, | \, \psi_l \rangle\ .
\ee
This Kac matrix and the resulting null-vector structure it predicts was previously studied in detail in \cite{Kac:1979fz}.

\paragraph{Level 1.} Let us compute the level-1 Kac matrix explicitly. There are three states with $J^3_0=m$, corresponding to the three vertex operators appearing in \eqref{eq:level-1 descendants}, namely
\be 
J^+_{-1} \ket{j,m-1}\ , \qquad J^3_{-1} \ket{j,m} \ , \qquad J^-_{-1} \ket{j,m+1}\ .
\ee
Their inner products are straightforwardly evaluated, and we have for example
\begin{align} 
\bra{1-j,m} (J_{-1}^3)^\dag J^+_{-1} \ket{j,m-1}&=\bra{1-j,m} [J_{1}^3, J^+_{-1}] \ket{j,m-1} \\
&=\bra{1-j,m} J^+_0 \ket{j,m-1} \\
&=m-1+j\ ,
\end{align}
where we have again used (\ref{eq:sl2action}). The other components can be worked out similarly, and the resulting Kac matrix is simply 
\be 
M_m^{(1)}=\begin{pmatrix} 
2m+k-2 & m-j & 0 \\ m+j-1 & -\frac{k}{2} & -m+j-1 \\ 0 & -m-j & -2m+k-2
\end{pmatrix} \label{eq:level-1 Kac matrix}
\ee
As a consistency check, notice that the determinant takes the simple form
\be 
\det M_m^{(1)}=2(k-2) \big(j-\tfrac{k}{2}\big)\big(j-1+\tfrac{k}{2}\big)\ ,
\ee
telling us that a level-1 null vector appears when $j=\frac{k}{2}$ or $j=1-\frac{k}{2}$, in accordance with the general formula \cite{Kac:1979fz}.  Note that $k=2$ is the critical level for $\mathfrak{sl}(2,\RR)$.

\subsection{The conformal block to leading and subleading order}\label{sec:confblockexp}
We can now assemble the ingredients we have discussed so far to determine the four-point function from a sum over three-point conformal blocks. 

\paragraph{General conformal block expansion.} Conformal blocks may be understood as summing over all descendants of an intermediate primary state, and hence reduce the sum over {\em all} intermediate states in, say, a four-point function to a sum over only the primary states. More explicitly, let us consider a four-point function of the form
\be 
F(x;z)=\big\langle V_{j_1,h_1}^{w_1}(0;0)V_{j_2,h_2}^{w_2}(1;1)V_{j_3,h_3}^{w_3}(\infty;\infty)V_{j_4,h_4}^{w_4}(x;z) \big \rangle\ ,
\ee
where we have used the global Ward identities to fix the locations of three of the vertex operators to their standard locations. (We also suppress right-moving quantities for conciseness.) Let us assume that $|x| < 1$ and $|z| < 1$. The idea of the conformal block expansion is to insert a complete set of states into this four-point function, i.e.\ to sum over an orthonormal basis (ONB) of the intermediate space of states
\begin{align}
    F(x;z)=\sum_{\ket{\psi} \text{ ONB of }\mathcal{H}} \big\langle V_{j_2,h_2}^{w_2}(1;1)V_{j_3,h_3}^{w_3}(\infty;\infty)  \, | \psi \rangle \langle \psi |\,  V_{j_1,h_1}^{w_1}(0;0)V_{j_4,h_4}^{w_4}(x;z)\big \rangle\ . \label{eq:naive conformal block insertion}
\end{align}  
Let us remember that ket states $\ket{\psi}$ can be represented by vertex operators inserted at $(x;z)=(0;0)$, while the bra state $\bra{\psi}$ corresponds to a vertex operator inserted at $(x;z)=(\infty;\infty)$. Thus, both factors in \eqref{eq:naive conformal block insertion} are three-point functions. By applying a scaling transformation to the first three-point function, we have
\be 
\langle \psi |\,  V_{j_1,h_1}^{w_1}(0;0)V_{j_4,h_4}^{w_4}(x;z)\big \rangle=x^{-h_1-h_4+h_\psi} z^{-\Delta_1-\Delta_4+\Delta_\psi} \langle \psi |\,  V_{j_1,h_1}^{w_1}(0;0)V_{j_4,h_4}^{w_4}(1;1)\big \rangle\ ,
\ee
where $\Delta$ is the worldsheet conformal dimension 
\be
\Delta = \frac{\frac{1}{4} + p^2}{(k-2)} +N - w h + \frac{k}{4} w^2 \ ,  \label{eq:SL(2,R) worldsheet conformal weight}
\ee
with $N$ being the excitation number before spectral flow. (For the vertex operators associated to $j_1$ and $j_4$, $N=0$.) The sum in \eqref{eq:naive conformal block insertion} may now be split into the sum over primaries, and the sum over the descendants in a given representation. The sum over the descendants factors into the holomorphic and antiholomorphic sector, and the sum over the holomorphic descendants is then a conformal block appearing in the decomposition. Let us denote such a conformal block by $\mathcal{F}_j^w(x;z)$, where all the other external parameters $j_i$, $h_i$ and $w_i$, are left implicit. Because of the presence of the delta-function on the spins in the three-point function \eqref{eq:three-point function}, we are actually only interested in the special case $j=\frac{k}{2}-j_2-j_3=1-\frac{k}{2}+j_1+j_4$ and thus we will suppress also the label $j$ and assume that it takes this value. As we noticed before around eq.~\eqref{eq:pole}, $j$ lies in the physical spectrum precisely when $k=3$ and thus we will eventually restrict to this value.

The relevant conformal block we need to compute is therefore 
\begin{multline}
\mathcal{F}^w(x;z)=\sum_{\ket{\psi} \text{ ONB of }\mathcal{R}_{j}^w}x^{-h_1-h_4+h_\psi} z^{-\Delta_1-\Delta_4+\Delta_\psi}\big\langle V_{j_2,h_2}^{w_2}(1;1)V_{j_3,h_3}^{w_3}(\infty;\infty) \, | \psi \rangle_\text{hol} \\
\times \langle \psi |\, V_{j_1,h_1}^{w_1}(0;0)V_{j_4,h_4}^{w_4}(1;1)  \big \rangle_\text{hol}\ ,\label{eq:conformal block state definition}
\end{multline}
where $\mathcal{R}_{j}^w$ denotes the $w$-spectrally flowed spin $j$ representation, and $\langle \cdots \rangle_\text{hol}$ is the holomorphic part of \eqref{eq:three-point function}, without the structure constants which we take to be\footnote{Perhaps it would be more appropriate to insert an additional factor of $i$ in the delta-function, since this is a delta-function in the imaginary part of the argument. We will however suppress it, and similarly a corresponding factor of $i$ that one might insert in the measure of $j$ in \eqref{eq:four-point function conformal block expansion}.} 
\be 
C\begin{pmatrix}
    j_1 & j_2 & j_3 \\ w_1 & w_2 & w_3
\end{pmatrix}=|\hat\Pi|^{-k} \delta\big(j_1+j_2+j_3-\tfrac{k}{2}\big)\ ,
\ee
where $\hat{\Pi}$ was defined in eq.~(\ref{eq:hatPi}). For example, if all three states are primary we simply have
\be \label{eq:3inf}
\big\langle j,h\, |  V_{j_1,h_1}^{w_1}(0;0)V_{j_4,h_4}^{w_4}(1;1) \rangle_\text{hol}^{(w)}=(a_1^{[3]})^{\frac{k}{4}(w_1-1)-h_1}\, (a_4^{[3]})^{\frac{k}{4}(w_4-1)-h_4} \, a_\infty^{\frac{k}{4}(w-1)-h} \ , 
\ee
where  $a_j^{[3]}$, $j=1,4$ as well as $a_\infty$ refer to the covering map coefficients as determined by the covering map with three critical points.\footnote{We use the superscript $\!^{[3]}$ on the covering map coefficients of the three-punctured sphere to distinguish them from the covering map coefficients $a_i$ of the four-punctured sphere that appear in \eqref{eq:four-point function}. We also denote by $a_0$ the covering map parameter of the three-point function at $0$, where $\ket{\psi}$ is inserted and by $a_\infty$ the corresponding parameter where $\bra{\psi}$ is inserted.} Similarly, for the other three-point function we write 
\be \label{eq:3zero}
\big\langle  V_{j_2,h_2}^{w_2}(1;1)V_{j_3,h_3}^{w_3}(\infty;\infty) \,| j,h\rangle_\text{hol}^{(w)}=(a_2^{[3]})^{\frac{k}{4}(w_2-1)-h_1}\, (a_3^{[3]})^{\frac{k}{4}(w_3-1)-h_4} \, a_0^{\frac{k}{4}(w-1)-h} \ .
\ee
With these definitions, the full four-point function can be expressed through the conformal blocks as
\begin{align} 
F(x;z)&= \sum_w \int \text{d} j \  \frac{1}{\mathcal{N}(j,w)}C\begin{pmatrix} j_1 & j_4 & 1-j \\ w_1 & w_4 & w \end{pmatrix} C\begin{pmatrix} j_2 & j_3 & j \\ w_2 & w_3 & w \end{pmatrix} \, |\mathcal{F}_j^w(x;z)|^2 \ , \label{eq:four-point function conformal block expansion}
\end{align}
where $\mathcal{N}(j,w)$ is the two-point function normalisation appearing in \eqref{eq:two-point normalization}.
The $j$-integral localises on the value $j=\frac{k}{2}-j_2-j_3$, and leads to the momentum-conserving delta-function in \eqref{eq:four-point function}.

\paragraph{Non-orthogonal basis.} It is convenient to generalise \eqref{eq:conformal block state definition} to a summation over a non-orthogonal basis, as is also usually done in the Virasoro case. (In fact, the two-point function pairs spin $j$ and spin $1-j$, and is thus strictly speaking non-orthogonal anyway.) For this, let us split the sum over the states $\ket{\psi}$ into a sum over fixed spacetime conformal weight $h$, and fixed level $N$ above the ground state.\footnote{Here, we define the level in the spectrally flowed sense as in Section~\ref{subsec:Kac matrix}.} The two-point function is block diagonal as different spacetime weights $h$ and levels $N$ do not mix. For a given block, the matrix of inner products is precisely given by the Kac matrix $M_{m}^{(N)}$ for $m=h-\frac{kw}{2}$ that we discussed above. Thus, for the same choice of basis $\ket{\psi_1},\dots,\ket{\psi_{P_3(N)}}$ that was used for the calculation of the Kac matrix, we have
\begin{multline}
    \mathcal{F}^w(x;z)=\sum_{N \ge 0}\sum_{h} \sum_{i,l} x^{-h_1-h_4+h} z^{-\Delta_1-\Delta_4-\frac{j(j-1)}{k-2}+ N -h w+\frac{k w^2}{4}} \\
\times \big\langle V_{j_2,h_2}^{w_2}(1;1)V_{j_3,h_3}^{w_3}(\infty;\infty) \, | \psi_i \rangle_\text{hol}\, \big(M_{h-\frac{kw}{2}}^{(N)}\big)_{i,l}^{-1} \, \langle \psi_l |\, V_{j_1,h_1}^{w_1}(0;0)V_{j_4,h_4}^{w_4}(1;1)  \big \rangle_\text{hol}\ . \label{eq:conformal block state definition Kac matrix}
\end{multline}
For a given principal series representation of $\mathfrak{sl}(2,\RR)_k$, this sum runs over all the values of $h$ of the form $h \in \ZZ+\frac{kw}{2}+\alpha$ for some $\alpha \in \RR/\ZZ$. We will treat $\alpha$ somewhat imprecisely for now. As we shall discuss below in Section~\ref{subsec:Euclidean theory}, we should really be discussing the theory which is analytically continued in the target space signature to Euclidean $\mathrm{AdS}_3$, in which case $\alpha$ is not a parameter and everything works out properly. 
\paragraph{Convergence and localisation.} Let us examine the definition of \eqref{eq:conformal block state definition} in more detail. We are particularly interested in the convergence properties of the sum. While for Virasoro conformal blocks, the corresponding sum has a convergence radius of $z=1$ (since conformal blocks have their closest singularity near the origin at $z=1$), this is not the case for \eqref{eq:conformal block state definition}, since it also depends on $x$. Consider the partial sum over only the affine primary states of $\mathcal{R}_j^w$, i.e.\ the $N=0$ term in \eqref{eq:conformal block state definition Kac matrix}. Up to overall factors that are independent of $h$, the sum over $h$ is then of the form
\be 
\sum_{h}  x^h z^{-h w} (a_0 a_\infty)^{-h}\ . \label{eq:sum over affine primary states}
\ee
Since the sum is unbounded in both directions, this is clearly somewhat formal. The conformal block expansion \eqref{eq:conformal block state definition Kac matrix} is therefore only sensible in the double scaling limit
\be 
|z| \ll 1\ , \qquad |x| \ll 1\ , \qquad x z^{-w} \ \text{finite}\ . \label{eq:double scaling limit}
\ee
The first condition is necessary for the subleading levels in \eqref{eq:conformal block state definition Kac matrix} to be suppressed, while the condition that $x z^{-w}$ is finite guarantees that all states in the infinite sum \eqref{eq:sum over affine primary states} are of the same order; this then also forces $x$ to be small. Thus the conformal block expansion requires us to move the operators close together, \emph{both on the worldsheet and in spacetime}.

With this understanding, we declare that \eqref{eq:sum over affine primary states} evaluates to a formal delta-function, i.e.
\be 
\sum_{h}  x^h z^{-h w} (a_0 a_\infty)^{-h}\longrightarrow \frac{z}{w} \, \delta\Big(z-\Big(\frac{x}{a_0 a_\infty}\Big)^{\frac{1}{w}}\Big)\ , \label{eq:formal delta function replacement}
\ee
which when we combine it with the right-movers leads to the localising delta-function in the four-point function \eqref{eq:four-point function0}. Obviously, this step is slightly illegal, since the sum over $h$ is treated as a formal sum and the delta function on the right hand side is holomorphic. In fact, the full delta-function appearing in \eqref{eq:four-point function0} cannot be factorised in this way. This problem will again be resolved in the Euclidean setting, see Section~\ref{subsec:Euclidean theory}, but we will use \eqref{eq:formal delta function replacement} (and suitable derivatives of it) for now.

The normalisation $\frac{1}{w}$ perhaps requires some explanation. In this sum, $h$ runs over values $\ZZ+\frac{kw}{2}+\alpha=\ZZ+\alpha'$,
\begin{align}
    \sum_{h \in \ZZ+\alpha'} x^h z^{-h w} (a_0 a_\infty)^{-h}\longrightarrow \delta\Big(1- \frac{x}{a_0 a_\infty z^w}\Big)\ .
\end{align}
We can then pass to \eqref{eq:formal delta function replacement} and this leads to a Jacobian factor of $\frac{1}{w}$. 

We should however mention that there is an ambiguity in taking the power $\frac{1}{w}$ in \eqref{eq:formal delta function replacement}. When we combine left- and right-movers as in \eqref{eq:four-point function conformal block expansion}, we should also sum over the $w$ possible values of taking the $w$-th root in the delta function. This additional sum is the sum over the representation label $\alpha$ that is shared between left- and right-movers and leads to an additional factor of $w$ if we view the correlator as a function of the $z_i$'s. Thus the two-point function normalisation \eqref{eq:two-point normalization} is appropriate to cancel the overall $w$-dependence. This factor of $w$ in the normalisation disappears when we pass to the string-theoretic two-point function, see Section~\ref{subsec:bosonic string} below.

\subsection{Matching with the four-point function to subleading order}\label{sec:confblockexpsub}

\paragraph{Level 0.} It is now straightforward to evaluate \eqref{eq:conformal block state definition Kac matrix} to leading order, i.e.\ for $N=0$. This gives
\begin{align} 
\mathcal{F}^w_{N=0}(x;z)&=\prod_{i=1}^4 (a_{i}^{[3]})^{\frac{k}{4}(w_i-1)-h_i} z^{-\frac{(2j_1-k+2)(2j_4-k+2)}{2(k-2)}+\frac{k}{4}(1+w^2-w_1^2-w_4^2)-1+w_1 h_1+w_4 h_4}x^{-h_1-h_4} \nonumber\\
&\qquad\times \sum_h (xz^{-w})^{h} (a_0 a_\infty)^{\frac{k}{4}(w-1)-h} \big(1+\mathcal{O}(z)\big)\\
&=\frac{1}{w}z^{-\frac{(2j_1-k+2)(2j_4-k+2)}{2(k-2)}}\big(x z^{-w_1} a_1^{[3]}\big)^{\frac{k}{4}(w_1-1)-h_1}(a_2^{[3]})^{\frac{k}{4}(w_2-1)-h_2}(a_3^{[3]})^{\frac{k}{4}(w_3-1)-h_3}   \nonumber\\
&\qquad \times \big(x z^{-w_4} a_4^{[3]}\big)^{\frac{k}{4}(w_4-1)-h_4}\big(z^{w+1} a_0 a_\infty \big)^{\frac{k}{4}(w+1-w_1-w_4)} \nonumber\\
&\qquad\times \delta\Big(z-\Big(\frac{x}{a_0 a_\infty}\Big)^{\frac{1}{w}}\Big)\big(1+\mathcal{O}(z)\big)\ .
\end{align}
We recall that this expression is valid in the double scaling limit \eqref{eq:double scaling limit}.
We claim that to this order, we can write this also as
\begin{multline} 
\mathcal{F}^w_{N=0}(x;z)=\frac{1}{w}z^{-\frac{(2j_1-k+2)(2j_4-k+2)}{2(k-2)}}(1-z)^{-\frac{(2j_2-k+2)(2j_4-k+2)}{2(k-2)}}\prod_{i=1}^4 a_i^{\frac{k}{4}(w_i-1)-h_i} \\
\times   \Big(\frac{\hat{\Pi}}{\hat{\Pi}_0^{[3]} \hat{\Pi}_\infty^{[3]}}\Big)^{-\frac{k}{2}} \delta(z-z_\gamma)\big(1+\mathcal{O}(z)\big)\ , \label{eq:exact conformal block}
\end{multline}
where $\hat{\Pi}_0^{[3]}$ and $\hat{\Pi}_\infty^{[3]}$ denote the factors associated to (\ref{eq:3zero}) and (\ref{eq:3inf}), respectively.
Here, $\gamma$ is the unique covering map ramified over the four points $0$, $1$, $\infty$ and $z$ which behaves as $x=\mathcal{O}(z_\gamma^w)$ as $x \to 0$.  
To check this, we need various identities on the behaviour of the coefficients of the covering map as $z \to 0$. To this order, we can check these identities using the formalism employed in \cite{Roumpedakis:2018tdb, Dei:2019iym}, since the relevant identities are the same as those needed to check the correct factorisation properties of correlation functions in a symmetric orbifold, but since we are not aware of a simple way to check the relevant identities at the next order, we simply verified them in \texttt{Mathematica} for a large number of cases. They correspond to the leading term in the equations \eqref{eq:xgamma identity} -- \eqref{eq:Pi identity} below.  Upon insertion of \eqref{eq:exact conformal block} into the conformal block expansion, we recover \eqref{eq:four-point function0} to leading order.

\paragraph{Level 1.} We determined all the ingredients to evaluate \eqref{eq:conformal block state definition Kac matrix} also for $N=1$, see \eqref{eq:level-1 J3 correlator} and \eqref{eq:level-1 Kac matrix}. Up to the $N=1$ order, the conformal block reads
\begin{align}
    \mathcal{F}^w_{N=1}(x;z)&=z^{\frac{k(w^2-w_1^2-w_4^2+1)}{4}-\frac{(2j_1-k+2)(2j_4-k+2)}{2(k-2)}-1+w_1 h_1+w_4 h_4}x^{-h_1-h_4} \prod_{i=1}^4 (a_i^{[3]})^{\frac{k}{4}(w_i-1)-h_i}\nonumber\\
    &\qquad\times \sum_{h} z^{-w h}x^h \big(a_0a_\infty\big)^{\frac{k}{4}(w-1)-h}\bigg[1+z \bigg(\frac{(2j_2-k+2)(2j_4-k+2)}{2(k-2)}\nonumber\\
    &\qquad\qquad+(h-w^{-1})\Big(\frac{2b_0 b_\infty}{a_0 a_\infty w}-\frac{b_0}{a_0}-\frac{b_\infty}{a_\infty}\Big)+\frac{b_0 (h_1w_1-h_4 w_4)}{a_0 w} \nonumber\\
    &\qquad\qquad +\frac{b_\infty (h_3w_3-h_2 w_2)}{a_\infty w}+\frac{k}{6}\Big(\frac{b_0(w^2-w_1^2+w_4^2-1)}{a_0w}\nonumber\\
    &\qquad\qquad+\frac{b_\infty(w^2+w_2^2-w_3^2-1)}{a_\infty w}-\frac{b_0 b_\infty (w^2-1)}{a_0a_\infty w^2}\Big)\bigg)+\mathcal{O}(z^2)\bigg]\ . \label{eq:conformal block N=1 level}
\end{align}
While this looks perhaps relatively complicated, many terms have cancelled out. In particular, notice that the $j_i$-dependence has become very simple and agrees to this order with the expected free boson dependence in \eqref{eq:four-point function0},
\be 
z^{-\frac{(2j_1-k+2)(2j_4-k+2)}{2(k-2)}}(1-z)^{-\frac{(2j_2-k+2)(2j_4-k+2)}{2(k-2)}}\ .
\ee
It is also remarkable that \eqref{eq:conformal block N=1 level} does not feature terms quadratic in $h$, even though these would naively be present from the definition \eqref{eq:conformal block state definition Kac matrix}. The terms linear in $h$ formally produce a derivative of the delta-function as in \eqref{eq:formal delta function replacement}, which can be viewed as coming from the expansion of $\delta(z-z_\gamma)$ to first order, since $z_\gamma$ is a non-trivial function of $x$, which receives corrections at subleading orders. In fact, \eqref{eq:conformal block N=1 level} still matches with \eqref{eq:exact conformal block}, even at this subleading order. One can confirm this explicitly using the identities 
\begin{subequations}
\begin{align}
    x_\gamma&=z^w a_0 a_\infty \bigg[1- \Big(\frac{2b_0 b_\infty}{a_0 a_\infty w}-\frac{b_0}{a_0}-\frac{b_\infty}{a_\infty}\Big)z+\mathcal{O}(z^2) \bigg]\ ,  \label{eq:xgamma identity}\\
    a_1&=z^{-w_1}x_\gamma a_{1}^{[3]}\Big(1-\frac{b_0 w_1}{a_0 w} z+\mathcal{O}(z^2)\Big)\ ,  \label{eq:a1 identity}\\
    a_2&=a_{2}^{[3]}\Big(1+\frac{b_\infty w_2}{a_\infty w} z+\mathcal{O}(z^2)\Big)\ , \label{eq:a2 identity}\\
    a_3&=a_{3}^{[3]}\Big(1-\frac{b_\infty w_3}{a_\infty w} z+\mathcal{O}(z^2)\Big)\ , \label{eq:a3 identity}\\
    a_4&=z^{-w_4} x_\gamma a_{4}^{[3]}\Big(1+\frac{b_0 w_4}{a_0 w} z+\mathcal{O}(z^2)\Big)\ , \label{eq:a4 identity}\\
    \hat{\Pi}&=\pm z^{\frac{1}{2}(w+1)(w_1+w_4-w-1)}\hat{\Pi}_0^{[3]} \hat{\Pi}_\infty^{[3]} (a_0 a_\infty)^{\frac{1}{2}(w_1+w_4-w-1)}\nonumber\\
        &\qquad\times\bigg[1+\bigg(\frac{ (2-2 w (w+3)-(w_1-3) w_1+(w_4-3) w_4+3 w (w_1+w_4))b_0}{6 w a_0}\nonumber\\
        &\qquad\qquad+\frac{(2-2 w (w+3)+(w_2-3) w_2-(w_3-3) w_3+3 w (w_1+w_4)+2) b_{\infty }}{6 w a_{\infty }}\nonumber\\
        &\qquad\qquad+\frac{ (w^2-3 (w_1+w_4-2) w-1) b_0 b_{\infty }}{3 w^2 a_0 a_{\infty }}\bigg)z+\mathcal{O}(z^2)\bigg]\ . \label{eq:Pi identity}
\end{align} \label{eq:covering map identities}%
\end{subequations}
The $\pm$ sign in \eqref{eq:Pi identity} depends on the choices of spectral flows. It cancels out once we combine left and right movers and we will thus suppress it in the following.
We found these identities by requiring the matching with \eqref{eq:exact conformal block} and confirmed them numerically for all covering maps with $\sum_i w_i \le 14$ (2052 instances). The \texttt{mathematica} code is attached for completeness as an ancillary file.

\paragraph{Crossing symmetry.} With \eqref{eq:exact conformal block} at hand, it is then simple to verify that we obtain \eqref{eq:four-point function0} when we insert \eqref{eq:exact conformal block} into \eqref{eq:four-point function conformal block expansion}. Therefore we demonstrated crossing symmetry up to this order.

\subsection{The Euclidean theory} \label{subsec:Euclidean theory}
Before explaining the argument demonstrating the equality of the conformal block with \eqref{eq:exact conformal block} to all orders, it is useful to first explain how Wick rotation in target space removes the subtleties with the delta-function that we encountered above.

\paragraph{Zero modes and $\SL(2,\CC)$ representations.} We treated the affine ground states $\ket{j,m}^{(w)}$ as a representation of the spectrally flowed zero modes $J_w^+$, $J_0-\frac{kw}{2}$ and $J^-_{-w}$, see eq.~(\ref{eq:specflowaction}), which form an $\mathfrak{sl}(2,\mathbb{R})$ algebra. However, the $x$-dependence of the correlator accounts for the transformation properties under the unflowed zero modes $J_0^+$, $J_0^3$ and $J_0^-$. In the Lorentzian theory, $J^a_0$ and the right-moving currents $\bar{J}_0^a$ are not related by complex conjugation and correspondingly, the parameter $x$ should be chosen to be real, with the right-moving analogue $\bar{x}$ also being real (and independent). However, for the delta-function localising correlators \eqref{eq:proposal npt function}, we clearly want to allow $x$ to be complex which means that we Wick rotated the zero modes $J_0^a$ and $\bar{J}_0^a$ into an $\mathfrak{sl}(2,\CC)$ algebra.
For consistency, the spectrally flowed modes $J_w^+$, $J_0-\frac{kw}{2}$ and $J^-_{-w}$, as well as their right-movers then also have to form an $\mathfrak{sl}(2,\CC)$ algebra. 
Thus, in the Euclidean theory, we should not have the states $\ket{j,m,\bar{m}}^{(w)}$ transforming as $\mathfrak{sl}(2,\RR)$ representations with respect to both left- and right-moving zero modes, but rather as a single $\mathfrak{sl}(2,\CC)$ representation. However, for $\mathfrak{sl}(2,\CC)$ representations, we cannot diagonalise both $J_0^3$ and $\bar{J}_0^3$ with discrete spectrum, since the Cartan torus of $\SL(2,\CC)$ has one compact and one non-compact direction. Rather than employing a basis in which $J_0^3$ and $\bar{J}_0^3$ are diagonalised, we should therefore trade the variables $m$ and $\bar{m}$ (or equivalently $h$ and $\bar{h}$) with a complex variable $y \in \CC$ such that\footnote{With respect to \cite{Dei:2021xgh, Dei:2021yom}, we replaced $j_i \to 1-j_i$. This is merely a convention and in any case irrelevant as long as we only consider the principal continuous series with $j_i \in \frac{1}{2}+i \RR$.}
\begin{subequations}
\begin{align} 
J_w^+ \ket{j,y}^{(w)}&=\partial_y \ket{j,y}^{(w)}\ , \\
\Big(J_0^3-\frac{kw}{2}\Big) \ket{j,y}^{(w)}&=\big(1-j+y \partial_y\big) \ket{j,y}^{(w)}\ , \\ \qquad J_{-w}^- \ket{j,y}^{(w)}&=\big(2(1-j)y+y^2 \partial_y\big) \ket{j,y}^{(w)}\ ,
\end{align} \label{eq:action zero modes y basis}%
\end{subequations}
and similar actions with $\bar{y}$ for the right-moving modes. 
\paragraph{Vertex operators.} Vertex operators in the Euclidean theory thus naturally depend on \emph{three} coordinates $V_j^w(x;y;z)$ and are formally obtained from the Lorentzian vertex operators by the replacement 
\be 
\sum_{h,\bar{h}} V_{j,h,\bar{h}}^w(x;z)\,  y^{h-1+j-\frac{kw}{2}} \bar{y}^{\bar{h}-1+j-\frac{kw}{2}} \longrightarrow V_j^w(x;y;z)\ .
\ee
One can check that the action of the spectrally flowed zero modes on $V_{j,h,\bar{h}}^w(x;z)$ correctly translate to the spectrally flowed zero modes \eqref{eq:action zero modes y basis}.
The structure constants in the Euclidean theory then simply take the form 
\begin{multline} 
\Big\langle \prod_{i=1}^3 V_{j_i}^{w_i}(x_i;y_i;z_i) \Big \rangle=\delta\big(j_1+j_2+j_3-\tfrac{k}{2})| \prod_{i<\ell} |z_i-z_\ell|^{-\frac{(2j_i-k+2)(2j_\ell-k+2)}{k-2}}\\
\times \Big|\hat\Pi^{-\frac{k}{2}}\prod_{i=1}^3 a_i^{j_i-\frac{k}{4}(w_i+1)} \Big|^2\delta^{2}(y_i-a_i)\ .
\end{multline}
\paragraph{Inner product.} We can repeat the previous discussion in this `$y$-basis'. The only qualitative change is the treatment of the sum over $h$ in the conformal block expansion \eqref{eq:conformal block state definition Kac matrix}. $y$ parametrises the full $\mathfrak{sl}(2,\CC)$ representation space and the gluing of the two three-point functions is achieved by taking the integral over $y$. To see what the correct measure for this is, consider the two-point function in the $y$-basis, determined in \cite{Dei:2021xgh},\footnote{There is another potential term when $j_1=j_2$, but we only consider the case $j_1=1-j_2$ here.}
\be 
\big \langle V_{j}^{w}(0;y_1;0)V_{1-j}^{w}(\infty;y_2;\infty) \big \rangle'\propto  |y_1|^{4j-2} \, \delta^2(1-y_1y_2)\propto|y_1|^{4j} \delta^2(y_1-y_2^{-1})\ ,
\ee
where the ${}'$ means that we omitted the momentum-conserving delta-function. The $y$-dependence of the two-point function is determined by the Ward identities, and we can view this part as the `holomorphic' piece as in the three-point function. There can still be a prefactor depending on $j$ and $w$ that keeps track of the normalisation of the two-point function.
We should thus insert $y^{-1}$ in the bra state, and multiply by $|y|^{-4j}$ to obtain an invariant pairing. Thus the conformal block is to leading order given by
\begin{multline}
|\mathcal{F}^w(x;z)|^2=\int \text{d}^2 y\  \big|y^{-2j}\langle  V_{j_2}^{w_2}(1;y_2;1)V_{j_3}^{w_3}(\infty;y_3;\infty) | j,y \big \rangle^{(w)}_\text{hol}\\
\times {}^{(w)}\!\langle 1-j,y^{-1} |\,  V_{j_1}^{w_1}(0;y_1;0)V_{j_4}^{w_4}(x;y_4;z)\big \rangle_\text{hol}
\big|^2 + \cdots \ .
\end{multline}
Notice that the integral over $y$ prevents the conformal block from factorising into a holomorphic and an anti-holomorphic part.
The `holomorphic part' of the three-point function is obtained from \eqref{eq:3inf} by going to the $y$-basis and using that
\be 
\sum_{h,\bar{h}} \big|a^{\frac{k}{4}(w-1)-h}y^{h-1+j-\frac{kw}{2}}\big|^2\longrightarrow\big|a^{j-\frac{k}{4}(w+1)}\big|^2 \delta^2(y-a)\ .
\ee
We can again move $x$ and $z$ to the locations $(x;z)=(1;1)$ by translating the scaling transformation to the $y$-basis. This leads to an overall power of $x$ and $z$.  Furthermore, the momentum conservation becomes now, see eq.~(\ref{eq:pole})
\be
j_1 + j_4 + 1-j = \tfrac{k}{2} \qquad \Longrightarrow \qquad  j = j_1 + j_4 + 1 - \tfrac{k}{2} \ . 
\ee
Combining all of this, we then obtain to leading order 
\begin{align}
    |\mathcal{F}^w(x;z)|^2&=\int \text{d}^2 y\  
    \Big|y^{-2j}x^{\frac{k}{2}(w-w_1-w_4-1)+2(j_1+j_4)-1} \nonumber\\
&\qquad\times z^{-\frac{(2j_1-k+2)(2j_4-k+2)}{2(k-2)}+\frac{k}{4}(w_1^2+w_4^2-w^2+2w+1)+(w+w_1)(1-j_1)+(w+w_4)(1-j_4)-3w-1} \nonumber\\
&\qquad\times\prod_{i=1}^4 (a_i^{[3]})^{j_i-\frac{k}{4}(w_i+1)} a_0^{j_1+j_4+1-\frac{k}{4}(w+3)} a_\infty^{-j_1-j_4-\frac{k}{4}(w-1)}\Big|^2 \nonumber\\
&\qquad\times \delta^2(x^{-1} y_1 z^{w_1} -a_1^{[3]})\, \delta^2(y_2-a_2^{[3]})\, \delta^2(y_3-a_3^{[3]}) \nonumber\\
&\qquad\times\delta^2(x^{-1}y_4 z^{w_4}-a_4^{[3]})\, \delta^2(y-a_0)\, \delta^2(x y^{-1} z^{-w}-a_\infty)\label{eq:conformal block y basis leading order 1st line}\\
&=\frac{1}{w^2}\int \text{d}^2 y\ \Big|z^{-\frac{(2j_1-k+2)(2j_4-k+2)}{2(k-2)}} (a_0 a_\infty x^{-1} z^w)^{-j_1-j_4-\frac{k}{4}(2w-w_1-w_4) - 1} \nonumber\\
&\qquad\times (z^{w+1} a_0 a_\infty)^{\frac{k}{4}(w+1-w_1-w_4)} \prod_{i=1}^4 a_i^{j_i-\frac{k}{4}(w_i+1)}\Big|^2 \nonumber\\
&\qquad \times \prod_{i=1}^4 \delta^2(y_i-a_i) \, \delta^2(y-a_0) \, \delta^2\Big(z-\Big(\frac{x}{a_\infty y}\Big)^{\frac{1}{w}}\Big) \label{eq:conformal block y basis leading order 2nd line}\\
&=\frac{1}{w^2} \Big| z^{-\frac{(2j_1-k+2)(2j_4-k+2)}{2(k-2)}}\Big|^2  \delta^2(z-z_\gamma)\nonumber\\
&\qquad\times \Big|\Big(\frac{\hat{\Pi}}{\hat{\Pi}_0^{[3]}\hat{\Pi}_\infty^{[3]}}\Big)^{-\frac{k}{2}}\prod_{i=1}^4   a_i^{j_i-\frac{k}{4}(w_i+1)}\Big|^2\delta^2(y_i-a_i) \ . \label{eq:conformal block y basis leading order}
\end{align}
In going from \eqref{eq:conformal block y basis leading order 1st line} to \eqref{eq:conformal block y basis leading order 2nd line} we have used the relations between the three-point and  four-point covering map constants, see eqs.~\eqref{eq:a1 identity}--\eqref{eq:a4 identity}, to bring the delta-functions into a more standard form. We have also inserted the localising value of $y=a_0$ into the prefactor $y^{-2j}$. Finally, to arrive at eq.~\eqref{eq:conformal block y basis leading order}, we integrated out $y$ and used the relations \eqref{eq:xgamma identity} and \eqref{eq:Pi identity} to rewrite the result in terms of the covering map data of the four-punctured sphere.
This is precisely what we wanted. In particular, the delta-function comes out on the nose.
One can similarly carry out the analysis at subleading order in the conformal block expansion and finds agreement, just as in the Lorentzian theory.
\subsection{A general argument} \label{subsec:general argument}
We now give a general argument that the conformal block in the Euclidean $y$-basis is given \emph{exactly} by
\begin{multline} 
|\mathcal{F}^w(x;z)|^2=\frac{1}{w^2} \Big| z^{-\frac{(2j_1-k+2)(2j_4-k+2)}{2(k-2)}}(1-z)^{-\frac{(2j_2-k+2)(2j_4-k+2)}{2(k-2)}}\Big(\frac{\hat{\Pi}}{\hat{\Pi}_0^{[3]}\hat{\Pi}_\infty^{[3]}}\Big)^{-\frac{k}{2}}\Big|^2 \\
\times \delta^2(z-z_\gamma)\prod_{i=1}^4   \Big|a_i^{j_i-\frac{k}{4}(w_i+1)}\Big|^2\delta^2(y_i-a_i) \ . \label{eq:exact block}
\end{multline}
This can be done by using the fact that conformal blocks can also be understood as a complete basis to the Ward identities of the theory. Thus, we can systematically impose the Ward identities on the four-point function and identify the conformal blocks obtained in this manner with the conformal blocks obtained by summing over intermediate states. 

\paragraph{Ward identities.} In general, there are two types of Ward identities: the recursion relations (that become differential equations in $y_i$, as well as the spacetime cross ratio $x$ after passing to the Euclidean theory) and the Knizhnik-Zamolodchikov equation (that is a differential equation in all $y_i$'s as well as the two cross-ratios $x$ and $z$). 
These constraints are discussed in detail in \cite{Dei:2021yom}.
It is possible to solve the recursion relations in closed form which expresses the solution in terms of a function depending on two variables. The solution takes the form
\begin{multline} 
\mathcal{F}^w(x;z)=|X_\emptyset|^{2(4-k-\sum_i j_i)} |X_{12}|^{2(j_1+j_2-j_3+j_4-2)} |X_{13}|^{2(j_1-j_2+j_3-j_4)} |X_{23}|^{2(-j_1+j_2+j_3-j_4)}\\
 \times |X_{34}|^{4j_4-4} \mathcal{F}\Big(\frac{X_{23}X_{14}}{X_{12}X_{34}};z\Big)\ . \label{eq:conformal block general w solution}
\end{multline}
Explicit expressions for the quantities $X_\emptyset$ and $X_{ij}$ can be found in \cite{Dei:2021yom}. They are polynomials of degree 1 in each $y_i$ whose $i$ appears in the subscript, and more complicated functions in $x$ and $z$.
It was also shown in \cite{Dei:2021yom} that the Knizhnik-Zamolodchikov equations applied to the spectrally flowed block $\mathcal{F}^w(x;z)$ reduce to the usual (unflowed) Knizhnik-Zamolodchikov equations for the function $\mathcal{F}(c;z)$.\footnote{For our conventions, it satisfies the unflowed KZ-equations with flipped spins $j_i \to 1-j_i$.} For special choices of the spins, $\mathcal{F}^w(x;z)$ satisfies additional Ward identities due to the appearance of null-vectors in the corresponding affine representations. These null-vector constraints also simply correspond to the null-vector constraints in the unflowed sector for the remaining function $\mathcal{F}(c;z)$. Since these constraints characterise conformal blocks completely, this means that the remaining function $\mathcal{F}(x;z)$ has to be the unflowed current algebra block (up to overall normalisation), which has been widely studied, see e.g.\ \cite{Teschner:1997ft,Teschner:1999ug}.

\paragraph{Localising blocks.} So far, we have discussed the generic blocks. For our purposes we would, however, like to specialise to the case $\sum_i j_i=k-1$ and $j=\frac{k}{2}-j_2-j_3$. Then the conformal block can be dramatically simplified. The reason for this simplification is that the solution can be completely localised on a codimension 5 sublocus in the parameter space $(y_1,y_2,y_3,y_4,x,z)$. This is rather involved to spell out and we refer to \cite{Dei:2022pkr} for details. A simpler route is to consider the spectrally flowed conformal block with external spins $1-j_i$, which satisfy  $\sum_i (1-j_i)=5-k$, i.e.\ to consider the conformal block related by reflection symmetry. In this case, the solution is only localised on a codimension one sublocus in the parameter space. This is because the first term in \eqref{eq:conformal block general w solution} reads $|X_\emptyset|^{-2}$. For a negative integer exponent, there is then an alternative solution given by replacing $|X_\emptyset|^{-2} \to \delta^2(X_\emptyset)$, since they obey the same differential equations. 

The condition $X_\emptyset=0$ is equivalent to imposing $z=z_\gamma$ for some covering map $\gamma$. We thus pick out one covering map and trade the delta-function $\delta^2(X_\emptyset)$ for $\delta^2(z-z_\gamma)$ at the cost of a Jacobian. Once we put $z=z_\gamma$ in the remaining factors $X_{ij}$, they simplify also dramatically and factorise into $(1-a_i^{-1} y_i)(1-a_j^{-1} y_j)$, up to an overall $y_i$-independent prefactor; the relevant identities can be found in \cite[Appendix C]{Dei:2022pkr}. In any case, after the dust settles, we find that the conformal block for the continuous series representation is, up to an overall normalisation that is independent of $(y_1,y_2,y_3,y_4,x,z)$, given by (compare with \cite[eq.~(3.15)]{Dei:2022pkr})
\begin{multline}
\mathcal{F}^w(x;z)\propto \delta^2(z-z_\gamma) \, |z|^{2(-j_2-j_3+\frac{k}{2})} \, |1-z|^{2(-j_1-j_3+\frac{k}{2})} \\
 \times \Big|\hat{\Pi}^{-\frac{k}{2}}\prod_{i=1}^4 a_i^{-\frac{k}{4}(w_i+1)-j_i}(1- a_i^{-1} y_i )^{-2j_i}\Big|^2\, \mathcal{F}_{j_\text{int}=\frac{k}{2}-1+j}(z;z)\ , \label{eq:reflected conformal block}
\end{multline}
where $\mathcal{F}(z;z)$ is the unflowed conformal block evaluated at $x=z$. Here, the internal spin of the unflowed conformal block is related to $j$ as $j_\text{int}=\frac{k}{2}-(1-j)$. The reflection comes about because we use different conventions to \cite{Dei:2021yom} here. More specifically, the redefinition $j \to \frac{k}{2}-j$ arises because in the three-point functions, one of the spins has to be redefined as $j \to \frac{k}{2}-j$ \cite{Dei:2021xgh} to relate the spectrally flowed three-point functions to the unflowed ones. One can also check this assignment explicitly by taking the small $z$ limit of \eqref{eq:reflected conformal block} and checking that this internal $j$ leads to the correct leading singularity. The external spins are given by $j_i$ since we flipped the spin twice, once because of our conventions and once because we are considering the spectrally flowed block with external states $1-j_i$. We can now reflect back to $j_i \to 1-j_i$, which is implemented via the shadow transform. In our case, the relevant integral reads
\be 
\int \text{d}^2 y' \ |y-y'|^{4j-4} \big|1-a^{-1} y' \big|^{-4j}\ .
\ee
This is simply the result of taking the shadow transform of one of the operators in a CFT two-point function, which is well-known to evaluate to a delta-function. If we formally perform this integral we obtain $\delta^2(a-y) |a|^{4j}$, up to an overall prefactor.

\paragraph{Unflowed block.} It remains to determine the unflowed conformal block $\mathcal{F}(z;z)$ at the coincidence limit $x \to z$ that appears in \eqref{eq:reflected conformal block}. This coincidence limit was studied in detail in \cite{Ponsot:2002cp}, see also \cite{Dei:2022pkr}. There is a singular and a regular branch of the conformal block in this limit. Since we want to evaluate $\mathcal{F}(z;z)$, we pick the regular part, and it is given by
\begin{align}
    \mathcal{F}_{j_\text{int}=\frac{k}{2}-1+j}(z;z)=\frac{\Gamma(2-2j)\Gamma(k-j_1-j_2-j_3-j_4)}{\Gamma(1-j_1-j_4-j+\frac{k}{2})\Gamma(1-j_2-j_3-j+\frac{k}{2})}  \mathcal{F}_{\alpha=b j_\text{int}}^\text{Vir}(\alpha_i=b j_i;z)\, . \label{eq:coincidence limit SL(2,R) block}
\end{align}
The block that appears on the right hand side is the standard Virasoro block with $\alpha_i$ the Liouville momenta. The central charge is given by $c=1=6(b+b^{-1})^2$ where $b^2=\frac{1}{k-2}$. This remarkable correspondence of $\mathrm{SL}(2,\RR)$ current algebra blocks and Virasoro conformal blocks is a special case of the $H_3^+$-Liouville correspondence \cite{Ribault:2005wp}.

Granting \eqref{eq:coincidence limit SL(2,R) block}, it is now simple to work out $\mathcal{F}(z;z)$ even more explicitly in the case where $\sum_i j_i=k-1$ and $j=\frac{k}{2}-j_2-j_3$, which corresponds exactly to the saturation of the background charge in the Virasoro language and momentum conservation,
\be 
\sum_i \alpha_i=Q=b+b^{-1}\ , \qquad \alpha=\alpha_1+\alpha_4\ .
\ee
This means that the conformal block can be evaluated via the Coulomb gas formalism \cite{Dotsenko:1984nm} where it corresponds to the leading order in conformal perturbation theory and is thus simply a free boson correlator,
\be 
\mathcal{F}(z;z)=|z|^{-4 \alpha_1\alpha_4} |1-z|^{-4\alpha_2 \alpha_4}=|z|^{-\frac{4j_1 j_4}{k-2}} |1-z|^{-\frac{4j_2 j_4}{k-2}}\ .
\ee
\paragraph{Normalisation.} Thus, the conformal block in the $y$-basis for $\sum_i j_i=k-1$ reads, up to overall normalisation
\begin{multline} 
\mathcal{F}^w(x;z)\propto \delta^2(z-z_\gamma) \, |z|^{-\frac{(2j_1-k+2)(2j_4-k+2)}{k-2}} \, |1-z|^{-\frac{(2j_2-k+2)(2j_4-k+2)}{k-2}} \\
 \times \Big|\hat{\Pi}^{-\frac{k}{2}}\prod_{i=1}^4 a_i^{-\frac{k}{4}(w_i+1)+j_i}\Big|^2\, \delta^2(y_i-a_i)\ .
\end{multline}
Since this result is obtained by solving Ward identities, it fixes the exact dependence of the conformal block on the variables $(y_1,y_2,y_3,y_4,x,z)$, but not yet its overall normalisation. For that, we can compare with the leading term in the conformal block expansion that we determined above \eqref{eq:conformal block y basis leading order}. This finally gives \eqref{eq:exact block} as the exact answer for the conformal block.

\paragraph{Crossing symmetry.} It is now trivial to see that the exact conformal block \eqref{eq:exact block} leads to the four-point function \eqref{eq:four-point function0} appropriately translated to $y$-space. In particular, it is crossing symmetric.

\section{String theory and holography}\label{sec:holography}
We now turn to the holographic application of our discussion, which we already sketched in the Introduction.
\subsection{Bosonic string on \texorpdfstring{$\mathrm{AdS}_3 \times {\rm X}$}{AdS3 x X}} \label{subsec:bosonic string}
Let us start with bosonic strings on a background $\mathrm{AdS}_3 \times {\rm X}$ at $k=3$. We take ${\rm X}$ to be described by a unitary CFT with $c_{\rm X}=26-\frac{3k}{k-2}=17$, so that the full worldsheet theory is critical. We can also assume that ${\rm X}$ has a discrete spectrum, so that the resulting string spectrum is at least discrete in the short string sector, but this will, strictly speaking, not be necessary.

We have a choice for the sigma model describing the $\mathrm{AdS}_3$ factor: we can take it to be the ordinary $\mathrm{SL}(2,\mathbb{R})_{k=3}$ WZW model or the CFT with the modified spectrum and correlators $\mathrm{SL}(2,\mathbb{R})_{k=3}'$ that we described above. Since both are consistent CFTs and have the $\mathrm{AdS}_3$ isometries as their symmetry group, they are equally good candidates to provide a sigma model description of $\mathrm{AdS}_3$ at the string scale. Moreover, the no-ghost theorem for $\mathrm{AdS}_3$ backgrounds \cite{Petropoulos:1989fc, Dixon:1989cg, Hwang:1990aq, Evans:1998qu, Maldacena:2000hw} carries over to the $\mathrm{AdS}_3$ background described by $\mathrm{SL}(2,\mathbb{R})_{k=3}'$ since the spectrum is more restricted for $\mathrm{SL}(2,\mathbb{R})_{k=3}'$. Let us thus consider the modified string background $\mathrm{AdS}_3' \times {\rm X}$, where $\mathrm{AdS}_3'$ is described by the $\SL(2,\RR)_{k=3}'$ WZW model.

\paragraph{Operator matching.} The string spectrum of this background coincides precisely with the single-particle spectrum of the symmetric orbifold $\text{Sym}^N(\mathbb{R} \times {\rm X})$ \cite{Eberhardt:2019qcl}, where $\mathbb{R}$ denotes a free boson factor. Notice that
\be 
c_{\mathbb{R} \times {\rm X}}=1+17=18\ ,
\ee
which coincides with the expectation $6k$ from the construction of the DDF operators realising the spacetime Virasoro algebra \cite{Giveon:1998ns}.
The matching of the string spectrum is essentially the observation that the mass-shell condition of the string coincides with the formula for the conformal weights of the symmetric orbifold \cite{Eberhardt:2019qcl}. Let $V_{\rm X}$ be an arbitrary vertex operator for the theory ${\rm X}$ of conformal dimension $\Delta_{\rm X}$. The worldsheet conformal weight of a level-$N$ descendant of the primary vertex operator $V_{p,h,\bar{h}}^w(x;z) V_{\rm X}(z)$ is, see eq.~\eqref{eq:SL(2,R) worldsheet conformal weight}
\be 
\Delta =\frac{1}{4} + p^2 +N - w h + \frac{3}{4} w^2 +\Delta_{\rm X} \overset{!}{=} 1\ . \label{eq:mass shell condition}
\ee
Solving for $h$ gives
\begin{align}
    h=\frac{3(w^2-1)}{4w}+\frac{\Delta_{\rm X}+p^2+N}{w}\ . \label{eq:spacetime conformal weight on shell vertex operator}
\end{align}
$\frac{3(w^2-1)}{4w}$ is the ground state conformal dimension of the twist-$w$ sector in the symmetric orbifold CFT of central charge $18$, $\Delta_{\rm X}+p^2+N$ is the spacetime conformal weight of the corresponding state in the untwisted sector, and the additional $\frac{1}{w}$ corresponds to the fact that modes of the symmetric orbifold are fractionally moded in the twisted sector. Finally, the orbifold projection corresponds to the fact that $h-\bar{h} \in \ZZ$ in the worldsheet theory. One can confirm this matching of operators also more systematically by deriving the worldsheet partition function and imposing physical state conditions or directly by evaluating the torus partition function in thermal $\mathrm{AdS}_3$ as was done for $\mathrm{AdS}_3 \times \mathrm{S}^3 \times \mathbb{T}^4$ in \cite{Eberhardt:2018ouy, Eberhardt:2020bgq}. This procedure strictly speaking gives the symmetric orbifold in the grand canonical ensemble, meaning $\bigoplus_{N=0}^\infty \mathrm{Sym}^N(\mathbb{R} \times {\rm X})$ with a chemical potential corresponding to $N$ kept fixed, which in turn can be identified with the string coupling, see \cite{Kim:2015gak, Eberhardt:2020bgq, Eberhardt:2021jvj, Aharony:2024fid} for further discussions.

We should mention that bosonic string theory naively also has a tachyon in the spectrum, coming from the continuous unflowed (i.e.\ $w=0$) representation with $N=0$. Indeed, provided that $\Delta_{\rm X}\le \frac{3}{4}$, as is for example the case for the vacuum state of ${\rm X}$, \eqref{eq:mass shell condition} can be solved, leading to 
\be
p =\pm \sqrt{\frac{3}{4}-\Delta_{\rm X}} \ .
\ee
This is not a highest weight representation in the dual CFT; in fact its energy is unbounded from below (since we can choose $h \in \RR$ arbitrarily). The existence of this sector would clearly be puzzling from the perspective of the dual CFT. However, as we showed, the worldsheet OPEs close on the $w \ge 1$ representations and thus these tachyonic states (that arise from the $w=0$ sector) can be consistently removed. The result is a tachyon-free bosonic string theory with a large number of degrees of freedom! We will comment further on this remarkable fact in the Conclusions, see Section~\ref{sec:concl}. 

\paragraph{Correlators.} It now follows straightforwardly from our previous discussion that sphere correlation functions of the primary vertex operators $V_{p,h,\bar{h}}^w(x;z) V_{\rm X}(z)$ match with the dual CFT. If we take the correlators \eqref{eq:proposal npt function} and integrate them over the moduli, we get
\begin{multline}
    \int \text{d}^2 z_4 \cdots \text{d}^2 z_n \ \left\langle \prod_{i=1}^n V_{p_i, h_i,\bar{h}_i}^{w_i}(x_i;z_i)V_{\mathrm{X},i}(z_i) \right\rangle_{\rm new}\!\!\!\!\! =\delta\Big(\sum\nolimits_i p_i \Big)\sum_{\gamma}   \prod_{i<j}|z_i^\gamma-z_j^\gamma|^{4p_ip_j} \\
    \times \Big|\hat\Pi^{-\frac{3}{2}} 
 \prod_{i=1}^n  a_i^{\frac{3(w_i-1)}{4}-h_i}\Big|^2 \left\langle \prod_{i=1}^n V_{\mathrm{X},i}(z_i^\gamma) \right\rangle\ , \label{eq:correlator matching}
\end{multline}
where $h_i$ is determined through \eqref{eq:spacetime conformal weight on shell vertex operator}. This coincides precisely with the large $N$ contribution to the correlation function in the symmetric orbifold $\text{Sym}^N(\RR \times {\rm X})$ as evaluated using the covering space method \cite{Lunin:2000yv, Pakman:2009zz, Dei:2019iym}. Note that the factors 
\be
\Big|\hat\Pi^{-\frac{3}{2}}  \prod_{i=1}^n  a_i^{\frac{3(w_i-1)}{4}-h_i}\Big|^2
\ee
are the conformal factors arising from lifting the correlation function to the appropriate covering space, while the remaining part of the correlator is simply the correlation function on the covering space (which is also a sphere). In that sense, the duality again follows the slogan suggested in \cite{Eberhardt:2019ywk} that the worldsheet \emph{is} the covering space appearing in the symmetric orbifold. This matching also extends to descendants on the worldsheet. An abstract argument for this comes from the existence of DDF operators which implement the action of worldsheet spacetime modes on the worldsheet. For $k>3$, these operators are built on top of the representation $j=1$ in the unflowed sector \cite{Kutasov:1999xu}. For $k=3$, this representation is not consistent with the Maldacena-Ooguri bound $\frac{1}{2}<j<\frac{k-1}{2}$ and the corresponding representation moves to the $w=1$ sector using the identification 
\be
\mathcal{D}^+_j=\mathcal{D}^{-,(w=1)}_{\frac{k}{2}-j}
\ee
of discrete representations of $\SL(2,\RR)_k$. For $k=3$, we hence precisely land on the $j=\frac{1}{2}$ representation in the spectrally flowed sector, which contains the vacuum state of the dual CFT. One then simply constructs on the worldsheet the vertex operator corresponding to the stress tensor in the dual CFT and extracts the appropriate modes. This then allows one to reduce the calculation of descendant correlators to those involving only primaries, see also \cite{Dei:2019osr, Bertle:2020sgd}, where this was verified explicitly in some examples.

Let us mention that \eqref{eq:correlator matching} also works for two-point functions and in fact the normalisation \eqref{eq:two-point normalization}. The procedure to do this was explained in \cite{Maldacena:2001km, Erbin:2019uiz}. Modding out by the residual M\"obius symmetry on the worldsheet has the effect of cancelling the delta-function $\delta(\Delta-\Delta')$ on the worldsheet (that is superfluous because it is already implied by the mass shell condition). Thus, we should translate the delta-function in \eqref{eq:two-point normalization} to a delta-function in the worldsheet weight and remove it to pass to the spacetime two-point function. The corresponding Jacobian obtained from \eqref{eq:SL(2,R) worldsheet conformal weight} compensates for the factor $\frac{1}{w}$ and indeed leads to canonically normalised spacetime two-point functions.

\paragraph{Marginal operator.} A more complicated duality involving the standard $\mathrm{AdS}_3$ sigma-model $\SL(2,\RR)_{k=3}$ was proposed in \cite{Eberhardt:2021vsx}. In that case, the dual CFT is a marginal deformation of the symmetric orbifold $\text{Sym}^N(\RR \times {\rm X})$ by a non-normalisable marginal operator from the twist-2 sector carrying momentum $p=\pm \frac{i}{2}$, so that \eqref{eq:spacetime conformal weight on shell vertex operator} is indeed unity.\footnote{Since the ground state energy of the twist-2 sector is $h=\frac{18}{24}(2-\frac{1}{2})=\frac{9}{8}>1$, there are no normalisable marginal operators in the twisted sector.} Since the marginal operator carries non-trivial momentum, turning it on will lead to correlators without a  momentum-conserving delta function. Instead, the delta functions get smeared out to poles in the full interacting theory, similar to what happens when constructing Liouville theory out of a linear dilaton theory \cite{Dotsenko:1984nm, Zamolodchikov:1995aa}. In \cite{Eberhardt:2021vsx, Dei:2022pkr}, it was verified in conformal perturbation theory that this indeed recovers the much more involved correlation functions \eqref{eq:three point function integral expression} of the $\SL(2,\RR)_{k=3}$ WZW model.

From the dual CFT perspective, the theory without the marginal deformation is trivially consistent, which leads to further indirect evidence that also the worldsheet theory that we have constructed is consistent, and the formula \eqref{eq:proposal npt function} is in fact valid for any $n$, even though we did not verify this on the worldsheet. We should also mention that turning on the marginal operator order-by-order was called the `near-boundary approximation' in \cite{Knighton:2023mhq}, and a direct derivation of worldsheet correlators in the Wakimoto representation was given in \cite{Hikida:2023jyc, Knighton:2023mhq, Knighton:2024qxd, Knighton:2024pqh}. We find it satisfying that our construction is purely axiomatic and does not rely on free field constructions, which are notoriously subtle. We also stress that the undeformed theory on its own is not consistent away from $k=3$, since the OPEs do not close on the set of physical vertex operators.

\subsection{Superstrings on \texorpdfstring{$\mathrm{AdS}_3 \times {\rm X}$}{AdS3 x X}}\label{sec:susy}

We expect similar considerations to also apply for the superstring, but since this involves a careful treatment of the GSO projection on the worldsheet, it is not entirely straightforward to give a very generic discussion. Instead we will describe in the following the situation for the three maximally supersymmetric backgrounds, involving, in turn, ${\rm X}= {\rm S}^3 \times {\rm S}^3 \times {\rm S}^1$, ${\rm X}= {\rm S}^3 \times \mathbb{T}^4$, and  ${\rm X}= {\rm S}^3 \times {\rm K3}$. 

\paragraph{Superstrings on $\mathrm{AdS}_3 \times {\rm S}^3 \times {\rm S}^3 \times {\rm S}^1$.} This case was already discussed in some detail in \cite{Gaberdiel:2024dva}, see also \cite{Gaberdiel:2018rqv,Giribet:2018ada}, and in fact, this example was part of the original motivation for our paper. If we take the $\mathfrak{sl}(2,\RR)^{(1)}$ factor to be at $k_{\rm s}=1$, the corresponding $\mathfrak{su}(2)^{(1)}$ algebras have $k^\pm_{\rm s}=2$, and thus, after decoupling the fermions, the worldsheet theory is described by 
\be
\mathfrak{sl}(2,\RR)_3 \oplus \mathfrak{su}(2)_0 \oplus \mathfrak{su}(2)_0 \oplus \hbox{$10$ fermions and the boson of ${\rm S}^1$} \ . 
\ee
The bosonic level $k^\pm=0$ $\mathfrak{su}(2)$ theories are trivial, i.e.\ only consist of one state with conformal dimension zero, and hence can be ignored. If we restrict the representations of the $\mathfrak{sl}(2,\RR)_3$ algebra to be the continuous representations with $w\geq 1$, i.e.\ if we consider the new WZW model from above, the spacetime spectrum becomes that of the symmetric orbifold of 
\be\label{eq:S3S3S1}
{\rm Sym}_N \bigl( {\cal S}_0^2 \bigr) = {\rm Sym}_N \bigl( \hbox{8 free fermions + 2 free bosons} \bigr)  \ , 
\ee
without any deformation \cite{Gaberdiel:2024dva}. Here ${\cal S}_0 \cong (\hbox{4 free fermions + 1 free boson})$ is the minimal theory with large ${\cal N}=4$ superconformal symmetry. One of the two bosons in (\ref{eq:S3S3S1}) is the compact boson of the ${\rm S}^1$ factor, while the other one is the boson that arises from our modified ${\rm AdS}_3$ theory. Its correlators are simply those of a (non-compact) free boson, and thus both bosons appearing in (\ref{eq:S3S3S1}) are free fields.\footnote{The precise nature of the non-compact boson could not be deduced from the analysis of \cite{Gaberdiel:2024dva}, but this is now clear.}

While it is relatively straightforward to read off the bosonic part of eq.~(\ref{eq:S3S3S1}), the fermions require more care. In the NS-R worldsheet formalism, all the left-moving fermions, say, are either in the NS-sector or in the R-sector, and the same is also true for the right-movers. Furthermore, the worldsheet theory needs to be GSO-projected. If we spectrally flow the $3$ free fermions from the ${\rm AdS}_3$ factor --- this is natural, since the spacetime conformal dimension is to be identified with the coupled $J^3_0$ eigenvalue --- then the GSO projection depends on whether $w$ is even or odd, see e.g.\ \cite{Argurio:2000tb,Giribet:2007wp,Ferreira:2017pgt}. In the dual CFT $w$ becomes the length of the (single-cycle) twisted sector, and we deduce that in the $w$-cycle twisted sector the seed theory of (\ref{eq:S3S3S1}) needs to be projected as 
\begin{align}
\hbox{$w$ odd} \qquad & \frac{1}{2} \bigl(1+(-1)^F\bigr)_{\rm NS} + \frac{1}{2} \bigl(1+(-1)^F\bigr)_{\rm R}  \label{wodd}\\
\hbox{$w$ even} \qquad & \frac{1}{2} \bigl(1-(-1)^F\bigr)_{\rm NS} + \frac{1}{2} \bigl(1-(-1)^F\bigr)_{\rm R} \ . \label{weven} 
\end{align}
Since there are $8$ free fermions, one can use the abstruse identity to rewrite this as the unprojected NS sector (for $w$ odd), and the unprojected R sector (for $w$ even), and this is indeed required in order to reproduce the correct twisted sector of the symmetric orbifold, see \cite{Gaberdiel:2018rqv} for a careful discussion. 

While theories with large ${\cal N}=4$ superconformal symmetry are usually quite subtle, this is in some sense the most straightforward supersymmetric ${\rm AdS}_3/{\rm CFT}_2$ duality: it can be directly formulated in the NS-R formalism, without any negative level bosonic worldsheet factors. In the dual CFT, all its exactly marginal operators come from the untwisted sector, and hence do not modify the symmetric orbifold structure. As a consequence, the string spectrum does not contain any modulus that would switch on the tension; this is somewhat reminiscient of the `island' theories of  \cite{Dabholkar:1998kv}. 
We also note that this dual pair is much simpler than the case of $k_{\rm s}^+=1$ (for which $k_{\rm s}<1$), that was discussed in \cite{Eberhardt:2017pty, Eberhardt:2019niq}. Based on the spectrum, the dual CFT in that case is given by $\text{Sym}^N(\mathcal{S}_\kappa)$ with $\kappa=k_{\rm s}^--1$, but it is not yet understood how this duality works on the level of the correlators, since the NS-R description naively breaks down.

\paragraph{Superstrings on $\mathrm{AdS}_3 \times \mathrm{S}^3 \times \mathbb{T}^4$.}\label{sec:T4}
The situation for the case of $\mathrm{AdS}_3 \times \mathrm{S}^3 \times \mathbb{T}^4$ is a little bit more complicated, but one can again give a relatively straightforward description in the NS-R formalism. At level $k_{\rm s}=1$,  the worldsheet theory is described by 
\begin{align}
& \mathfrak{sl}(2,\RR)^{(1)}_1 \oplus \mathfrak{su}(2)^{(1)}_1  \oplus \mathbb{T}^4  \ \cong \ 
\mathfrak{sl}(2,\RR)_3 \oplus \mathfrak{su}(2)_{-1} \oplus  \hbox{$6$ free fermions}  \oplus \mathbb{T}^4\ ,
\end{align}
where in going to the right-hand-side we have decoupled the fermions in the usual manner.\footnote{The $\mathfrak{su}(2)_{-1}$ factor leads to non-unitarities even after passing to the string-theoretic BRST cohomology and we should further restrict the space of physical states as happens naturally in the hybrid formalism \cite{Eberhardt:2018ouy}. This will in particular also remove the continuum of states from $\mathrm{AdS}_3$. We will do this below directly in the spacetime theory.} (The $\mathbb{T}^4$ factor is supersymmetric, i.e.\ it contains both $4$ bosons and $4$ fermions.) If we take the $\mathfrak{sl}(2,\RR)_3$ factor to be described by our new CFT, the dual spacetime CFT is equal to the symmetric orbifold of 
\be\label{eq:dualCFT}
{\rm Sym}_N \Bigl( \RR \oplus \mathfrak{su}(2)_{-1} \oplus \hbox{$4$ fermions}  \oplus \mathbb{T}^4\Bigr) \ , 
\ee
where we have removed two of the worldsheet fermions as a consequence of the usual physical state condition. Next we rewrite 
\be\label{eq:sympbos}
\RR \oplus \mathfrak{su}(2)_{-1} \ \cong \ \hbox{$4$ symplectic bosons} \ , 
\ee
where we have used the free field realisation of e.g.\ \cite[Appendix~C.2]{Eberhardt:2019ywk}. In particular, it follows from the discussion in \cite[Section~2.1]{Dei:2020zui} that the extra boson that arises in addition to $\mathfrak{su}(2)_{-1}\cong \mathfrak{sl}(2,\RR)_1$ (and that is often denoted by $U$) is non-compact since it has a continuous spectrum.

On the other hand, the $4$ symplectic bosons can now also be interpreted as $4$ fermionic ghosts, and they therefore remove the $4$ fermions from (\ref{eq:dualCFT}). Thus we just produce the symmetric orbifold of $\mathbb{T}^4$. As regards the GSO projection, we need to evaluate the combination of (\ref{wodd}) and (\ref{weven}) for the full fermionic theory, before the free fermions have been removed by the symplectic bosons. (After all, the symplectic bosons are not affected by the GSO-projection.) Thus we can apply the same argument as above and rewrite the result in terms of $8$ unprojected fermions whose modes are half-integer moded for $w$ odd, and integer moded for $w$ even. If we take the $4$ symplectic bosons in (\ref{eq:sympbos}) to be similarly half-integer moded for odd spectral flow (and integer moded for even spectral flow), then they cancel precisely $4$ of the $8$ fermions in eq.~(\ref{wodd}) and (\ref{weven}), and we end up exactly with the spectrum of the symmetric orbifold of $\mathbb{T}^4$. 

\paragraph{Superstrings on $\mathrm{AdS}_3 \times \mathrm{S}^3 \times {\rm K3}$.}\label{sec:K3}
The situation for the case where $\mathbb{T}^4$ is replaced by K3 is similar, in particular, if we consider the case where ${\rm K3} = \mathbb{T}^4 / \mathbb{Z}_2$. Most of the analysis of the $\mathbb{T}^4$ case goes directly through; the only subtlety concerns the nature of the GSO projection in the $\mathbb{Z}_2$ twisted NS-NS sector of the orbifold where there are fermionic zero modes that are affected by the orbifold action. Before we impose the orbifold projection, there are $16$ ground states from the fermionic zero modes, and they contribute 
\be
(y + 2 + y^{-1}) \, (\bar{y} + 2 + \bar{y}^{-1}) 
\ee
to the partition function, where $y$ and $\bar{y}$ denote the chemical potentials with respect to the left- and right-moving $\mathfrak{u}(1)$ current. Before imposing the orbifold projection, the twisted sector partition function is therefore 
\be
2^4 \, q^{\frac{1}{4}} \, \bar{q}^{ \frac{1}{4}} \, (y + 2 + y^{-1}) \, (\bar{y} + 2 + \bar{y}^{-1}) \, 
\left| \prod_{n=1}^{\infty} \frac{(1 + y q^n)^2 ( 1 + y^{-1} q^n)^2}{(1 - q^{n-\frac{1}{2}})^4} \right|^2  \ ,
\ee
where the factor of $2^4$ comes from the fact that there are $16$ $\mathbb{Z}_2$ fixed points on the torus.

For type IIB we now choose the same GSO projection for the left- and right-movers, and hence, from the ground states, the states transforming as 
\be
\Bigl( ( y+y^{-1})  (\bar{y} + \bar{y}^{-1}) + 4 \Bigr) 
\ee
survive. In particular, they rive rise to $\frac{1}{2}$-BPS states, and the resulting partition function is the familiar supersymmetric K3 sigma model.\footnote{Note that specifying the action of the GSO-projection on the ground states then also fixes which excited states survive the orbifold and GSO-projection.} On the other hand, for type IIA we choose the opposite GSO projection for the left- and right-movers, and hence, from the ground states, the states transforming as 
\be
\Bigl(2 (y+y^{-1}) + 2 (\bar{y} + \bar{y}^{-1}) \Bigr) 
\ee
survive. While this also leads to a modular invariant partition function, there are now no $\frac{1}{2}$-BPS states from the twisted sector, and hence the resulting partition function does not describe the cohomology of K3. 

Thus we deduce that type IIB superstring theory on on $\mathrm{AdS}_3 \times \mathrm{S}^3 \times {\rm K3}$ is exactly dual to the symmetric orbifold of K3, while for the type IIA case, the dual CFT is also of the form 
\be
{\rm Sym}_N \Bigl( \mathbb{T}^4 / \mathbb{Z}_2 \Bigr) \ , 
\ee
but the orbifold acts differently in the twisted sector and hence the seed theory is not the usual K3 CFT (and in particular not supersymmetric).

\paragraph{Heterotic strings on $\mathrm{AdS}_3 \times \mathrm{S}^3 \times \mathbb{T}^4$.} 
Another interesting example\footnote{MRG thanks Kimyeong Lee for a useful discussion about this question.} is to consider heterotic strings ($\mathrm{E}_8 \times \mathrm{E}_8$ or $\text{Spin}(32)/\ZZ_2$) on $\mathrm{AdS}_3 \times \mathrm{S}^3 \times \mathbb{T}^4$ \cite{Kutasov:1998zh, Hohenegger:2008du}. We take the left-movers to be described by the supersymmetric theory and take them to be at $k_\text{s}=1$. After decoupling the free fermions, the worldsheet theory takes the form
\be 
\mathfrak{sl}(2,\mathbb{R})_3 \oplus \mathfrak{su}(2)_{-1} \oplus \text{10 chiral fermions} \oplus \mathbb{T}^4 \oplus \text{16 chiral bosons}\ ,
\ee
where the $\mathbb{T}^4$ is non-supersymmetric.
The presence of the $\mathfrak{su}(2)_{-1}$ means that the localisation argument of \cite{Eberhardt:2018ouy} applies, and we have no choice but to take $\mathfrak{sl}(2,\mathbb{R})_3$ to be the modified theory that we discussed in this paper.  $\mathfrak{su}(2)_{-1}$ is a non-unitary theory and contrary to the type II case, unitarity cannot be rescued with the fermions, because we only have left-moving fermions on the worldsheet. Thus, this background \emph{does not} satisfy the no-ghost theorem.
We can nevertheless continue and as in the type II case on $\mathbb{T}^4$, the spacetime theory is naively
\be 
\text{Sym}_N \bigl(\mathbb{R} \oplus \mathfrak{su}(2)_{-1} \oplus \text{8 chiral fermions} \oplus \mathbb{T}^4 \oplus \text{16 chiral bosons}\bigr)\ . 
\ee
As above, we can replace $\mathbb{R} \oplus \mathfrak{su}(2)_{-1}$ with four symplectic bosons, which cancel 4 of the chiral fermions. We then obtain the theory
\be 
\text{Sym}_N(\mathbb{T}^4 \oplus \text{4 chiral fermions} \oplus \text{4 chiral symplectic bosons} \oplus \text{16 chiral bosons})\ .
\ee
Thus, apart from the internal lattice of the 16 chiral bosons, the dual CFT is still given by the symmetric orbifold of a supersymmetric $\mathbb{T}^4$, but where the statistics of the right-moving fermions has been changed to bosons. In particular, the spin structure of the symplectic bosons is coupled to the spin structure of the fermions. As mentioned above, the no-ghost theorem does not hold and correspondingly this CFT is not unitary, but the discussion in this paper shows that both the spectrum and the correlators (at least at large $N$) will match.

\section{Conclusions}\label{sec:concl}
In this paper, we have constructed a consistent worldsheet theory with $\SL(2,\RR)_k$ symmetry at $k=3$, which is different from the standard WZW model. We demonstrated that the four-point functions as computed from the conformal block expansion of the three-point functions localise in moduli space and are crossing symmetric. The resulting correlators are vastly simpler than for generic levels, see eq.~\eqref{eq:proposal npt function}. In particular, this allowed us to show that bosonic string theory on $\mathrm{AdS}_3 \times {\rm X}$ (where the ${\rm AdS}_3$ factor on the worldsheet is described by this model) is exactly dual to the undeformed symmetric orbifold $\text{Sym}^N(\RR \times {\rm X})$. 

This is to be contrasted with the case where the worldsheet theory involves the standard $\SL(2,\RR)$ WZW,
for which the dual CFT must be deformed by a certain marginal operator in the twist-2 sector. Analogous statements also hold for the superstring, although it is difficult to make general statements (because of the GSO projection), and we focused on explaining how this fits with what is known about holography on the backgrounds $\mathrm{AdS}_3 \times \mathrm{S}^3 \times {\rm X}$ with ${\rm X}= \mathrm{S}^3 \times \mathrm{S}^1$, $\mathbb{T}^4$, and $\mathrm{K}3$. Let us now discuss a few interesting observations and open questions.

\paragraph{Consistent bosonic string theories.} Our construction gives a tachyon-free, consistent bosonic string theory. Only a handful of such theories are known, and they usually feature a very small number of physical degrees of freedom \cite{Witten:1991yr, Ginsparg:1993is, Collier:2023cyw, Collier:2024kmo}. The usual lore is that the \emph{effective} central charge of the matter worldsheet theory can be at most $c_\text{eff}=2$. The reason for this is that the effective central charge controls the Cardy growth of states, and hence the behaviour of the worldsheet partition functions near the  boundary of moduli space. Given that the \emph{effective} central charge of the $bc$ ghost system is $c_{\text{eff},\, bc}=-2$, the total effective central charge is $c_\text{eff}-2$, and the moduli space integral diverges for $c_\text{eff}>2$. This is the manifestation of the existence of the tachyon, and it tells us, in particular, that the theory has a non-vanishing tadpole, i.e.\ that we are expanding around the wrong vacuum.

Somewhat remarkably, our theory exposes a loophole in this lore. The moduli space integrals in our theory are not divergent because they are delta function localised with the support away from the boundaries of moduli space. In fact, the effective worldsheet central charge is $c_\text{eff}=3+17=20$, since $\mathrm{SL}(2,\RR)_{k=3}$ contributes three degrees of freedom. The theory is clearly consistent (even non-perturbatively) because $\text{Sym}^N(\mathbb{R} \times {\rm X})$ is. Admittedly, the background we constructed is very far from resembling semiclassical gravity, but it nevertheless shows that the landscape of consistent bosonic string theories is much bigger than one may have thought. 

\paragraph{The theory on higher genus surfaces.} We can in principle attempt to put the worldsheet theory on higher genus surfaces. In particular, it was shown in \cite{Eberhardt:2020akk} that the $\SL(2,\RR)$ Ward identities at $k=3$ also admit localising solutions to the possible covering maps at higher genus, see also \cite{Knighton:2020kuh} for the supersymmetric case. There are, however, no particularly useful expressions available for the higher genus contributions to the symmetric orbifold correlators, and thus the analysis is far less explicit. A formal proof for the agreement with the symmetric orbifold correlators was given in \cite{Hikida:2023jyc, Knighton:2024pqh}. From an axiomatic point of view, these higher genus correlators are, however, somewhat confusing. In particular, these arguments seem to suggest that the torus one-point and torus two-point functions vanish since all covering maps of the once- or twice-punctured sphere have genus 0. However, from the perspective of the conformal block expansion on the worldsheet, this cannot be true. Indeed, one can compute an $n$-point function on the torus by expanding around the degeneration where a once-punctured torus is glued to an $(n+1)$-punctured sphere. For example, in the case of $g=1$ and $n=2$, we could expand the correlator as follows,
\be 
    \begin{tikzpicture}[baseline={([yshift=-.5ex]current bounding box.center)}]
        \begin{scope}[xscale=.7, yscale=.7]
                    \draw[very thick, out=180, in=180, looseness=2] (-2,1.7) to (-2,-1.7);
                    \draw[very thick, bend right=40] (-2.9,0.1) to (-1.1,.1);
        		    \draw[very thick, bend left=40] (-2.75,0) to (-1.25,0);
        			\begin{scope}[shift={(2,1.2)}, xscale=.4]
            			\draw[very thick] (0,-.5) arc (-90:90:.5);
            			\draw[very thick, dashed] (0,.5) arc (90:270:.5);
        			\end{scope}
        			\begin{scope}[shift={(2,-1.2)}, xscale=.4]
            			\draw[very thick] (0,-.5) arc (-90:90:.5);
            			\draw[very thick, dashed] (0,.5) arc (90:270:.5);
        			\end{scope}
        			\begin{scope}[xscale=.4]
            			\draw[very thick, RoyalBlue] (0,-.9) arc (-90:90:.9);
            			\draw[very thick, dashed, RoyalBlue] (0,.9) arc (90:270:.9);
        			\end{scope}
        			\draw[very thick, out=0, in=180] (-2,1.7) to (0,.9) to (2,1.7);
        			\draw[very thick, out=0, in=180] (-2,-1.7) to (0,-.9) to (2,-1.7);
        			\draw[very thick, out=180, in=180, looseness=2.5] (2,-.7) to (2,.7);
           \node at (2.45,-1.2) {$1$};
           \node at (2.45,1.2) {$2$};
		\end{scope}
   \end{tikzpicture}\hspace{-.3cm}=\sum_{|\psi \rangle \text{ ONB}} 
          \begin{tikzpicture}[baseline={([yshift=-.5ex]current bounding box.center)}]
          \begin{scope}[xscale=.7, yscale=.7]
                    \draw[very thick, out=180, in=180, looseness=2] (-2,1.7) to (-2,-1.7);
                    \draw[very thick, bend right=40] (-2.9,0.1) to (-1.1,.1);
        		    \draw[very thick, bend left=40] (-2.75,0) to (-1.25,0);
        			\begin{scope}[xscale=.4]
            			\draw[very thick] (0,-.5) arc (-90:90:.5);
            			\draw[very thick, dashed] (0,.5) arc (90:270:.5);
        			\end{scope}
        			\draw[very thick, out=0, in=180] (-2,1.7) to (0,.5);
        			\draw[very thick, out=0, in=180] (-2,-1.7) to (0,-.5);
           \node at (0,1) {$|\psi \rangle$};
		\end{scope}
   \end{tikzpicture}
   \times 
       \begin{tikzpicture}[baseline={([yshift=-.5ex]current bounding box.center)}]
        \begin{scope}[xscale=.7, yscale=.7]
        			\begin{scope}[shift={(2,1.2)}, xscale=.4]
            			\draw[very thick] (0,-.5) arc (-90:90:.5);
            			\draw[very thick, dashed] (0,.5) arc (90:270:.5);
        			\end{scope}
        			\begin{scope}[shift={(2,-1.2)}, xscale=.4]
            			\draw[very thick] (0,-.5) arc (-90:90:.5);
            			\draw[very thick, dashed] (0,.5) arc (90:270:.5);
        			\end{scope}
        			\draw[very thick, out=0, in=180] (0,.5) to (2,1.7);
        			\draw[very thick, out=0, in=180] (0,-.5) to (2,-1.7);
        			\draw[very thick, out=180, in=180, looseness=2.5] (2,-.7) to (2,.7);
           \draw[very thick] (0,0) circle (.2 and .5);
           \node at (2.45,-1.2) {$1$};
           \node at (2.45,1.2) {$2$};
           \node at (0,1) {$\langle \psi|$};
		\end{scope}
   \end{tikzpicture}\ . \label{eq:genus 1 two-point}
\ee
Vanishing of the once-punctured torus would therefore also imply the vanishing of all such correlation functions, which clearly cannot be the case. Instead, we suspect that the one-point function of the identity, and the two-point functions also receive contributions from the torus, arising from correctly implementing the operator normalisations in the grand canonical ensemble \cite{Aharony:2024fid}. It would be interesting to understand this properly.

\section*{Acknowledgments}
We thank Alejandra Castro, Andrea Dei, Rajesh Gopakumar, Kimyeong Lee, Wei Li, Edward Mazenc, Sameer Murthy, Mukund Rangamani and Vit Sriprachyakul for useful discussions. 
LE is supported by the European Research Council (ERC) under the European Union’s Horizon 2020 research and innovation programme (grant agreement No 101115511).  The work of MRG is supported by a personal grant from the Swiss National Science Foundation, and the work of his group at ETH is also supported in part by the Simons Foundation grant 994306  (Simons Collaboration on Confinement and QCD Strings), as well as the NCCR SwissMAP that is also funded by the Swiss National Science Foundation. 

\appendix 

\section{Conventions}\label{app:Conventions}
\subsection{Symmetry algebra}
Before analytic continuation to Euclidean spacetime signature, the theory has an $\mathfrak{sl}(2,\mathbb{R})$ symmetry algebra. The $\mathfrak{sl}(2,\RR)_k$ commutation relations read
\begin{subequations}
\begin{align}
    [J_m^3,J_n^\pm]&= \pm J_{m+n}^\pm\ , \\
    [J_m^3,J_n^3]&=-\frac{k}{2} m \delta_{m+n,0}\ , \\
    [J_m^+,J_n^-]&=k m \delta_{m+n,0}-2J_{m+n}^3\ .
\end{align}
\end{subequations}
The Sugawara construction gives Virasoro generators with central charge $c=\frac{3k}{k-2}$.

\subsection{Vertex operators}\label{app:ConventionsOPEs}
We consider vertex operators in the principal series of $\mathfrak{sl}(2,\mathbb{R})$ with $\mathfrak{sl}(2,\mathbb{R})$ spin $j=\frac{1}{2}+i p$. The theory we will construct does not have any discrete series vertex operators. We will denote spectrally flowed affine primary vertex operators by $V_{j,h,\bar{h}}^w(x;z)$ with $j \in \frac{1}{2}+i \RR$ and $w \in \ZZ_{\ge 1}$. $(h,\bar{h})$ are the magnetic quantum numbers and label the different states in the ground state representation. The vertex operators have the following defining OPEs with the symmetry currents:
\begin{subequations} 
\begin{align}
J^+(\zeta) V_{j,h}^w(x;z)&=\frac{(J^+_{w} V_{j,h}^w)(x;z)}{(\zeta-z)^{w+1}}+\mathcal{O}((\zeta-z)^{-w})\ , \\
\big(J^3(\zeta)-x J^+(\zeta)\big) V_{j,h}^w(x;z)&=\frac{h V_{j,h}^w(x;z)}{\zeta-z}+\mathcal{O}(1)\ , \\
\big(J^-(\zeta)-2x J^3(\zeta)+x^2 J^+(\zeta)\big)  V_{j,h}^w(x;z)&=(\zeta-z)^{w-1} (J^-_{-w} V_{j,h}^w)(x;z)+\mathcal{O}((\zeta-z)^{w})\ .
\label{eq:regular-combination-OPE}
\end{align}\label{eq:x-OPEs}%
\end{subequations}
We have the identifications
\begin{subequations}
\begin{align}
(J_0^+ V_{j,h}^w)(x;z)&=\partial_x V_{j,h}(x;z)\ , \label{eq:J0p action}\\
(J_{\pm w}^+ V_{j,h}^w)(x;z)&=\left(h-\tfrac{kw}{2}\pm j\right) V_{j,h\pm 1}^w(x;z)\ .\label{eq:Jwp action}
\end{align} \label{eq:leading J action}%
\end{subequations}
These definitions follow from the properties of spectral flow as well as the definition of the $x$-coordinate. We have suppressed the $\bar{h}$ label here since it just goes along for the ride.

\subsection{Covering maps} \label{subapp:Conventionscoveringmaps}
We will frequently make use of expansion parameters of covering maps. We denote covering maps by $\gamma(z)$, where $\gamma:\mathbb{CP}^1 \to \mathbb{CP}^1$, ramified over some number of points $z_i$ with ramification indices $w_i$. 

\paragraph{Existence of covering maps.} The existence of covering maps imposes $n-3$ conditions on the data specified by $\{z_i,x_i,w_i\}_{i=1,\dots,n}$. Furthermore, the degree of the covering map is given by the Riemann-Hurwitz formula
\be \label{eq:Riemann Hurwitz}
N =  1 + \frac{1}{2}\sum_{i=1}^{n} (w_i-1) \ . 
\ee
In this paper, we shall be mostly interested in the case $n=3$ and $n=4$. For $n=3$, a unique covering map exists provided that the constraints $w_i\le N \in \ZZ$ are satisfied. These translate into triangle inequalities and a parity constraint
\be 
w_i \le w_j+w_k-1\ , \qquad \sum_i w_i \in 2\ZZ+1
\ee
for all choices of distinct $\{i,j,k\}$. For $n=4$, a covering map $\gamma$ exists provided that the cross ratios of the four-punctured sphere satisfy a polynomial relation. Up to normalisation, this polynomial was denoted by $P_{w_1,w_2,w_3,w_4}(x;z)$ in \cite{Dei:2021yom}. For example,
\be 
P_{3,2,3,2}(x;z) \propto x^2-9 x z^2+16 x z^3-9 x z^4+z^6\ . \label{eq:Hurwitz polynomial example}
\ee
The $z$-degree of this polynomial is known as the Hurwitz number in the mathematical literature and counts the number of inequalivalent covering maps of the four-punctured sphere with cross ratio $x$ and given ramification indices. Thus, different $z$-solutions are labelled by the covering map $\gamma$ and we denote them by $z_\gamma$, which we view as a function of $x$ (as well as the dicrete data $w_1,\dots,w_4$ and $\gamma$.) The $x$-degree of \eqref{eq:Hurwitz polynomial example} corresponds to the number of OPE channels that we will encounter in a given four-point function with these spectral flow indices. It is also convenient to denote by $x_\gamma$ the cross-ratio in $x$-space, which is a function of $z$.

\paragraph{Expansion parameters.} Let us first assume that none of the $z_i$'s or $x_i$'s is equal to $\infty$; the case when one $z_i$ and/or one $x_i$ equals $\infty$ will be discussed below. Near the ramification points $z_i$, the covering map takes the canonical form
\be \label{eq:covering map coefficients}
\gamma(t)=x_i+a_i(t-z_i)^{w_i}+ b_i (t-z_i)^{w_i+1} + \cdots\ , 
\ee
for some coefficients $a_i$ and $b_i$. By definition of the ramification structure, the first $w_i-1$ Taylor coefficients vanish. Since the covering map is completely specified by the collection $\{z_i,x_i,w_i\}_{i=1,\dots,n}$ (satisfying the constraints discussed above), we view the parameters $a_i$ and $b_i$ as implicit functions of this data. 

In the following it is convenient to fix $x_1=0$, $x_2=1$ and $x_3=\infty$, while $x_4=x$ is the cross ratio. We similarly set $z_1=0$, $z_2=1$, $z_3=\infty$ and $z_4=z$.
Extra care is needed when defining $a_3$ and $b_3$. We claim that we should set
\be 
\gamma(t)^{-1}=a_3 t^{-w_3}+b_3 t^{-w_3-1}+ \cdots\ .
\ee
The logic for this arises because $\gamma(t^{-1})^{-1}$ is the covering map that maps $0$ to $0$ with ramification index $w_3$. It is still correctly normalised since it maps $1$ to $1$.

One can also check that when we repeat the derivation of the recursion relations in Section~\ref{sec:recursion} with one field inserted at $x_3=z_3=\infty$ that one correctly obtains the solution \eqref{eq:recursion solution} with this definition of $a_3$. Thus, this is indeed the `correct' definition of $a_3$ for our purposes.

We also need the following quantity, defined by taking the product over all residues of the covering map, 
\be 
\Pi\equiv w_3^{-w_3-1}\prod_{a=1}^{N-w_3} \Pi_a\ ,\label{eq:Pi definition}
\ee
where
\be \label{eq:Pia definition}
\gamma(z) \sim \frac{\Pi_a}{z-\ell_a} 
\ee
near the poles $\ell_a$ of $\gamma$. There are $N-w_3$ such simple poles, where $N$ is the degree of the covering map, since $\gamma$ has by definition a pole of order $w_3$ at $t=\infty$. The inclusion of the factor $w_3^{-w_3-1}$ in the definition \eqref{eq:Pi definition} leads to symmetric formulae in all the fields. 

It is convenient to consider a renormalised version $\hat{\Pi}=\prod_{i=1}^n w_i^{\frac{1}{2}(w_i+1)} \Pi$ as in \eqref{eq:hatPi}, as this simplifies various formulae in the paper. This amounts to changing the normalisations of the vertex operators in a suitable way and is thus simply a matter of convenience. In particular, we have
\be 
\hat{\Pi}_{w,1,w,1}=1
\ee
with this definition. In the formulae \eqref{eq:four-point function}, we also make use of $\hat{\Pi}$ when none of the operators are inserted at $\infty$. In that case, we simply define it by applying a global conformal transformation to the previous definition.

\bibliographystyle{JHEP}
\bibliography{bib}
\end{document}